%% file: main_arxiv.tex
\documentclass[%
 reprint,
 superscriptaddress,
nofootinbib,
 amsmath,amssymb,
 aps,
 pre,
floatfix,
]{revtex4-2}

\usepackage{color}

\usepackage{algorithm}
\usepackage{algpseudocode}
\usepackage{graphicx}
\usepackage{dcolumn}
\usepackage{bm}
\usepackage{hyperref}
\usepackage[dvipsnames]{xcolor} 
\usepackage{mathtools}
\usepackage{ulem}  

\DeclareMathOperator*{\argmax}{arg\,max}
\begin{document}


\title{Learning strategies for optimised fitness in a model of cyclic dominance}

\author{Honghao Yu}
\affiliation{%
 Yusuf Hamied Department of Chemistry, University of Cambridge, Lensfield Road, Cambridge CB2 1EW, United Kingdom}%
\author{Robert L. Jack}%
\affiliation{%
 Yusuf Hamied Department of Chemistry, University of Cambridge, Lensfield Road, Cambridge CB2 1EW, United Kingdom}%
\affiliation{%
 Department of Applied Mathematics and Theoretical Physics, University of Cambridge, Wilberforce Road,
Cambridge CB3 0WA, United Kingdom}%

\normalem 

\date{\today}

\begin{abstract}
A major problem in evolutionary biology is how species learn and adapt under the constraint of environmental conditions and competition of other species. Models of cyclic dominance provide simplified settings in which such questions can be addressed using methods from theoretical physics. We investigate how a privileged (``smart'') species optimises its population by adopting advantageous strategies in one such model. We use a reinforcement learning algorithm, which successfully identifies optimal strategies based on a survival-of-the-weakest effect, including directional incentives to avoid predators. We also characterise the steady-state behaviour of the system in the presence of the smart species and compare with the symmetric case where all species are equivalent.
\end{abstract}

\maketitle

\section{Introduction}\label{sec:intro}
Ecological systems consist of large numbers of individuals, interacting through cooperation or competition, and surviving under complex environmental constraints such as limited resources and space.  As such, they are naturally studied by statistical mechanical models where populations of several (or many) species interact via competition or co-operation~\cite{lotka1910contribution, volterra1931variations, hofbauer1998evolutionary, frey2010evolutionary}.  An interesting class of these models involves three species with a relationship of cyclic dominance, analogous to the game of rock-paper-scissors~\cite{may1975nonlinear,kerr2002local,reichenbach2007mobility}.  This situation can be realised in experiments on \textit{E. coli}~\cite{kerr2002local, kerr2006local, hibbing2010bacterial, nahum2011evolution, nadell2016spatial}, and is also relevant more generally~\cite{jackson1975alleopathy, buss1979competitive, sinervo1996rock, gilg2003cyclic, kirkup2004antibiotic, lankau2007mutual,  curatolo2020cooperative, gude2020bacterial}.

Models of cyclic dominance support spiral patterns which arise from 
the combination of ``diffusion'' (individuals' motion) with local ``reactions'' (for example predation and reproduction) The spirals are formed by the species chasing one another, as may be generically expected in systems with non-reciprocal interactions~\cite{fruchart2021non, loos2023long, dinelli2023non, duan2023dynamical, chiacchio2023nonreciprocal, avni2023non}. Such patterns are also relevant in the biological setting~\cite{budrene1991complex, koch1994biological, nakamasu2009interactions, liu2011sequential, yamanaka2014vitro, barbier2018generic}.
The pattern formation has been studied in detail for simple models, focussing in particular on the case where the species all have equivalent behaviour, so that the system is invariant under their cyclic permutation~\cite{reichenbach2006coexistence, nahum2011evolution, reichenbach2007mobility, reichenbach2007noise, reichenbach2008self, peltomaki2008three, berr2009zero, dobramysl2018stochastic}.  Such models also support a fixation transition between a biodiverse state (with all three species present) and an absorbing (fixed) state where only one species survives~\cite{reichenbach2007mobility, reichenbach2007noise, hanson2012beyond, nadell2016spatial, hibbing2010bacterial}.  Particles' mobility plays a crucial role in this transition~\cite{reichenbach2007mobility, reichenbach2008instability, szczesny2014characterization}.

In the context of these
simplified ecological models it is also natural to consider how individuals or species can learn and optimize their behaviour~\cite{gerhard2021hunting, muinos2021reinforcement, falk2021learning, borra2022reinforcement}, or adapt to their environment~\cite{colabrese2017flow, bellemare2020autonomous, mandralis2021learning, monderkamp2022active, kaur2023adaptive}.  (This is the subject of evolutionary game theory~\cite{hofbauer1998evolutionary, traulsen2009exploration, traulsen2009stochastic, hindersin2019computation}.)  Even in simple systems with three cyclically dominating species, complex and counter-intuitive phenomenon can emerge.  For example, when three species have different predation rates, the species with the weakest predation tends to dominate: this counter-intuitive behaviour is referred to as the survival of the weakest~\cite{frean2001rock, berr2009zero}.  To address the complexity of spatial models,
reinforcement learning (RL) techniques~\cite{sutton2018reinforcement} are naturally applied to species optimization and learning~\cite{wang2020reinforcement, gerhard2021hunting, park2021co}, as well as being fruitfully exploited in more general non-equilibrium physical settings~\cite{colabrese2017flow, verma2018efficient, reddy2018glider, cichos2020machine, muinos2021reinforcement, chennakesavalu_probing_2021, vansaders2023informational}.

This work applies these ideas to a model of cyclic dominance.  Understanding how individual species survive and evolve is of fundamental interest to evolutionary biology~\cite{hofbauer1998evolutionary, nahum2011evolution}.  Starting from the model of~\cite{reichenbach2007mobility}, we introduce several new features, to arrive at a situation in which one privileged (``smart'') species seeks to increase its population, for which it faces a complex optimisation problem.     We address this via an RL scheme in which the species changes its behaviour incrementally, to adapt to its environment.  The modifications to~\cite{reichenbach2007mobility} include a natural (spontaneous) death process that acts on all species, and a hunger level for each particle, which provides an incentive for predation.  For the parameters that we consider, this means that species can only survive if their prey is also present, so the smart species must optimise its population while maintaining a biodiverse state. This aspect makes the optimisation problem challenging.  To solve it, the smart species can learn by adjusting its predation rate, and by adopting directional strategies that bias individuals' motion.  For example, they may choose to evade their predators, or hunt their prey.

The RL algorithm successfully improves the fitness of the smart species, by exploiting the survival-of-the-weakest effect.  (This effect is robust, despite the additional features of hunger and natural death in our model.)  We optimise the parameters of the smart species under two sets of external conditions which differ strongly in their total population densities, due to different natural death rates.
 In both cases, the smart species gains an advantage by evading its predators.  In the less dense case, it also benefits from spreading into empty regions.  We discuss the effects of these strategies on the pattern formation and spatial correlations, and we use these results to interpret the competitive advantage of the smart species.

This paper is organised as follows.  We describe the model definition and simulation methods in Sec.~\ref{sec:model}.  We present the phase behaviour of this model in Sec.~\ref{sec:phase_behaviour}.   We discuss the reinforcement learning algorithm in Sec.~\ref{sec:learning} and show its results in Sec.~\ref{sec:res-RL}.  Building on these results, Sec.~\ref{sec:result} explores in more detail the optimal strategies and the mechanisms by which the smart species increases its population.
We conclude our study in Sec.~\ref{sec:conclusion}.

\section{Model}\label{sec:model}

\subsection{Model Definition}\label{sec:model_definition}

\begin{figure}[!t]
    \centering 
    \includegraphics[width=0.48\textwidth]{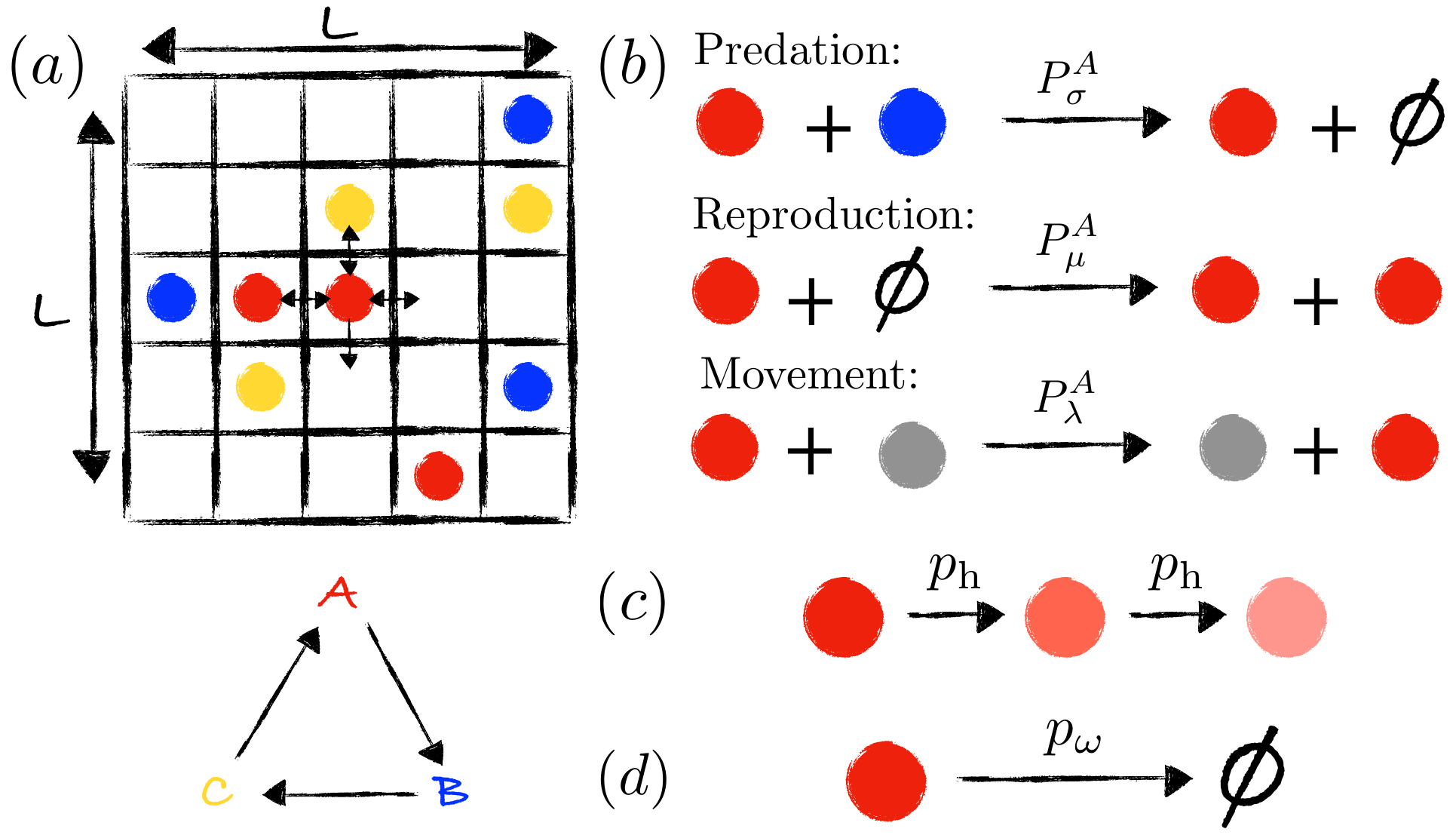}
    \caption{
    (a) Individual particles live on a two-dimensional lattice with volume exclusion and can interact with the four nearest neighbours.  Three species of particles exhibit cyclic dominance, where the arrows indicate predation.  
    (b) Dynamical rules for individual particles, using the red ($A$) species as an example.  Each particle can choose among three actions: predation, reproduction, and movement.
    (c) Individual particle increases hunger level over time.
    (d) Individual particle experiences natural death.  
    }
    \label{fig:model_definition}
\end{figure}

We consider a model with three species of particles ($A,B,C$)  on a two-dimensional square lattice of size $L\times L$ with periodic boundary conditions.   Each lattice site can be occupied by a single particle ($A, B, C$) or be vacant ($\varnothing$), see Fig.~\ref{fig:model_definition}(a). We adopt throughout this work the convention that species $A,B,C$ are coloured red, blue, and yellow, respectively.  The particle dynamics is a generalisation of the rock-paper-scissors (RPS) system of~\cite{reichenbach2007mobility}, and is also related to the May-Leonard model~\cite{may1975nonlinear}.  The species undergo predation, reproduction, and movement with rates $\sigma,\mu,\lambda$ respectively: these involve interactions with their nearest neighbours on the lattice, summarised as:
\begin{equation}\label{eq:actions-simple}
\begin{aligned}
    AB& \xrightarrow{\sigma_A} A\varnothing, 
    &&BC \xrightarrow{\sigma_B} B\varnothing,  
    & &CA \xrightarrow{\sigma_C} C\varnothing,
    \\
    A\varnothing& \xrightarrow{\mu_A} AA 
    &&B\varnothing \xrightarrow{\mu_B} BB, 
    & &C\varnothing \xrightarrow{\mu_C} CC,
    \\
    AS& \xrightarrow{\lambda_A} SA,
    &&BS \xrightarrow{\lambda_B} SB, 
    & &CS \xrightarrow{\lambda_C} SC,
\end{aligned}
\end{equation}
where $S$ may be any species, or an empty site.  This scheme allows for different species to have different rates for predation/reproduction/movement, for example, $\sigma_A,\sigma_B,\sigma_C$.
Note that the predation mechanism describes the cyclic domination among the three species: $A$ dominates (predates on) $B$, also $B$ dominates $C$, and $C$ dominates $A$.  
Due to the volume exclusion rule, reproduction can only be performed when the neighbouring site is empty.  
We use $X$ to denote a generic species and 
we write $\eta_X(\bm{r})$ for the number of particles of species $X$ at position $\bm{r}$.

In the following, we will allow a privileged (``smart'') species to adjust its predation and movement rates, to optimise its population.  To ensure that this optimisation problem captures the main challenges facing real species, we introduce two extra model features.  First, we introduce spontaneous death: each particle dies with rate $\omega$, independent of its species and its environment
\begin{equation}
    A \xrightarrow{\omega} \varnothing,
    \qquad
    B \xrightarrow{\omega} \varnothing,
    \qquad
    C \xrightarrow{\omega} \varnothing \; .
\label{eq:death}
\end{equation}
Second, we incorporate that particles must consume food in order to reproduce.  This is achieved by endowing each particle with a hunger level, with higher levels being the most hungry.  These are denoted by primes on the particle species: $A^0$ for level 0, $A^{\prime}$ for level 1, $A^{\prime\prime}$ for level 2.  Each particle increases its hunger stage with rate $h$: for species $A$ we have
\begin{equation}
\begin{aligned}
    A^0 \xrightarrow{h} A^{\prime} \xrightarrow{h} A^{\prime\prime},
\end{aligned}
\label{eq:hunger}
\end{equation}
with similar processes for species $B,C$.  (The parameter $h$ is the same for all species.)  A particle's hunger level is reset to zero when it undergoes a predation step and particles are born in a hungry state (see below for further details).  We take level 2 as the highest hunger level and we refer to particles in this level as \emph{hungry} particles.  These particles have a reduced reproduction rate 
\begin{equation}
\mu_{X}'' = {\cal H}\mu_X
\end{equation} 
where $\mathcal{H}$ is called the hunger reproduction factor and $X$ may be any of $A,B,C$.

There are obvious generalisations to include more hunger levels, and to have reproduction rates with more complicated dependence on the hunger level, or indeed to have other rates dependent on this level.  The key point for this work is that a system with only one species must converge to a state where all particles are hungry: if one has additionally $\mu'' < \omega$ then these particles die faster than they can reproduce, so the system will tend to an extinct state with no particles at all.  This ensures a non-trivial optimisation problem for the smart species, in that they can only survive as long as sufficient prey is available for them to eat.  (This situation is also more realistic in the ecological setting.)

\begin{figure*}
\begin{minipage}{\linewidth}
\begin{algorithm}[H]
  \caption{Discrete-time model dynamics}
  \label{alg:MC}
   \begin{algorithmic}[1]
   \State initialise each site independently according to Eq.~\eqref{eq:p_init}.
   \For{$t={1},...,T$} 
   \State update the hunger level all the particles in the system
   \For{$n={1},...,L\times L$} 
   \State choose a random lattice site $i$.
   \If{site $i$ is occupied by a particle}  
   \State particle dies with probability $p_{\omega}$.
   \If{the particle does not die}  
   \State choose random action ${\cal A}$ with probability $P_X({\cal A})$, according to Eq.~\eqref{eq:actions_prob}.
   \State choose random direction ${\cal D}$  with probability $P({\cal D})$, according to Eqs.~(\ref{eq:P0D},\ref{eq:P1D},\ref{eq:P2D}).
   \If{action ${\cal A}$ is allowed in direction ${\cal D}$}
   \State perform the action.
   \EndIf
   \EndIf
   \EndIf
   \EndFor
   \EndFor
   \end{algorithmic}
\end{algorithm}
\end{minipage}
\end{figure*}

To summarise these changes to the bare model of \eqref{eq:actions-simple}, we write 
\begin{equation}\begin{aligned}
    AB& \xrightarrow{\sigma_A} A^0\varnothing, 
    &&BC \xrightarrow{\sigma_B} B^0\varnothing,  
    & &CA \xrightarrow{\sigma_C} C^0\varnothing,
    \\
    A\varnothing& \xrightarrow{\mu_A} AA^{\prime\prime}, 
    &&B\varnothing \xrightarrow{\mu_B} BB^{\prime\prime}, 
    & &C\varnothing \xrightarrow{\mu_C} CC^{\prime\prime},
    \\
    AS& \xrightarrow{\lambda_A} SA,
    &&BS \xrightarrow{\lambda_B} SB, 
    & &CS \xrightarrow{\lambda_C} SC,
\end{aligned}\end{equation}
where it is now explicit that particles are born in the hungry state (for example $A^{\prime\prime}$), and their hunger level is set to zero by predation; the symbols $A,B,C$ denote particles that may be in any hunger level.  {(It is implicit here that the reproduction rate $\mu_X$ depends on the hunger level of the reproducing particle; we also explain in the following that the rates may be different according to particles' local environments.)}

\subsection{Formulation as discrete-time Markov process}
\label{sec:mc}

The model described so far can be used to define a Markov process in continuous time.  However, we take here a different route which is convenient for computer simulation: we define our model as a discrete-time Markov process, which we simulate by Monte Carlo (MC) dynamics.  The computational method is given as Algorithm~\ref{alg:MC}.   Particles may perform actions ${\cal A}$ taken from the set $\{\sigma,\mu,\lambda,\iota\}$ which respectively indicate selection, reproduction, and movement (as above), as well as remaining idle ($\iota$).  These actions also involve a randomly chosen neighbour denoted by ${\cal D}$ (direction) which is chosen from the set $\{{\rm left},{\rm right},{\rm up},{\rm down}\}$.

On each MC update, one chooses a random site $i$, a random action ${\cal A}$ and a random direction ${\cal D}$.  The site is chosen uniformly at random.  If the site is empty then nothing happens.  Otherwise, the particle on that site dies with probability 
\begin{equation}
p_\omega =  \omega \tau,
\label{eq:om-p-om}
\end{equation}
 where $\tau$ is the time step.  If there is no such death then an action is chosen according to the particle species $X$ as 
\begin{align}
    P_{X}({\sigma}) & = \tau \sigma_X ,
    & &
    P_{X^\ell}({\mu})   = \tau \mu_{X,\ell}  ,
    \nonumber \\ 
    P_{X}({\lambda}) & = \tau \lambda_X ,
    & &
    P_{X^\ell}({\iota})  = 1 - \tau (\sigma_X + \mu_{X,\ell} + \lambda_X)
\label{eq:actions_prob}
\end{align}
where $X^\ell$ indicates a particle with hunger level $\ell$, and $\mu_{X,\ell}$ is the corresponding reproduction rate [either $\mu_X$ or $\mu_X''$, see \eqref{eq:hunger}].  The time step $
\tau$ must be chosen small enough that $ P_{X^\ell}({\iota})\geq0$ for all species (and hunger levels).
In the simplest case, the direction ${\cal D}$ is also chosen randomly among the 4 neighbours, $P({\cal D})=(1/4)$, see however Sec.~\ref{sec:loca_para} below.  (The choice of direction is always independent of the action.) 
Given the random action and the random neighbour, it may be that the action is not allowed (for example, reproduction is only allowed if the neighbour is empty).  If the action is possible then it is performed.  The idle action ($\iota$) is always allowed; it leaves the system state the same.

We define an MC sweep (MCS) to be $L^2$ MC updates, so that each particle attempts on average one action per sweep.  In between each sweep, we increase the hunger level of every particle independently with probability $p_h=\tau h$.

To connect this process with a continuous-time formulation of the dynamics, one would take $\tau$ to be the time per MCS.  However, the relationship between continuous and discrete time formulations is not trivial here because (for example) each update involves either a death or another action (but not both).
 Throughout the following, we fix
$\sigma_B=\sigma_C=1$ and $\mu_A=\mu_B=\mu_C=1$, these rates serve as a baseline against which other rates can be compared (the choices of other parameters are discussed below).  Note that $\sigma_A$ is not fixed: this reflects that $A$ will be the smart species in the following, which may adjust its rates to optimise its population.  
When simulating the system, we report time in MCS.

\subsection{Further simulation details}
\label{sec:params-etc}

The model definition depends on several parameters.  Our main concern here is the effect of singling out a smart species that behaves differently from the others.  To explore this in a controlled way, we keep some of the parameters fixed.  In particular, we keep all parameters equal between species $B,C$, only adjusting the properties of the (smart) species $A$.  We also fix the reproduction rate of species $A$ equal to the other two (for example, we might imagine that this rate is fixed for the organism of interest, while the rates for predation and movement are behavioural choices and hence easier for the individuals to adjust).  We fix the parameters $p_h=0.02$, ${\cal H}=0.02$ associated with hunger levels.   Alternative values for these parameters would change quantitatively the model behaviour but we expect the qualitative results of this work to be robust.

Simulations are initialised by setting every site independently to be either empty or to a randomly chosen species, with probabilities
\begin{equation}
\label{eq:p_init}
p_{\rm init}(\varnothing) = 1/2 , \qquad p_{\rm init}(A) = p_{\rm init}(B) = p_{\rm init}(C) = 1/6
\end{equation}
All particles have initial hunger level 0.
  Lattice sizes are either $L=120$ or $L=300$, a comparison of the behaviour in these cases is useful for (qualitative) assessment of finite-size effects.  Note that we perform finite-size scaling with all parameters fixed, in contrast to \cite{reichenbach2007mobility} which took $\lambda \propto L^2$.
  
The population of a given species is measured by its number density $\rho_X=N_X/L^2$, where $N_X$ is the number of particles belonging to species $X$.   We write $\rho_{\rm total} = \rho_A + \rho_B + \rho_C$.
In addition to particles' species and hunger levels, we also follow several other statistics for each particle: their age (number of MCS since birth) and their predation/reproduction counts, which are the numbers of times they performed the predation and reproduction actions (number of prey consumed and number of children produced).  
We collect histograms of particle ages and predation/reproduction counts at their times of death, which may happen either spontaneously ($\omega$) or by predation ($\sigma$).

\begin{figure}[t!]
    \centering 
    \includegraphics[width=0.45\textwidth]{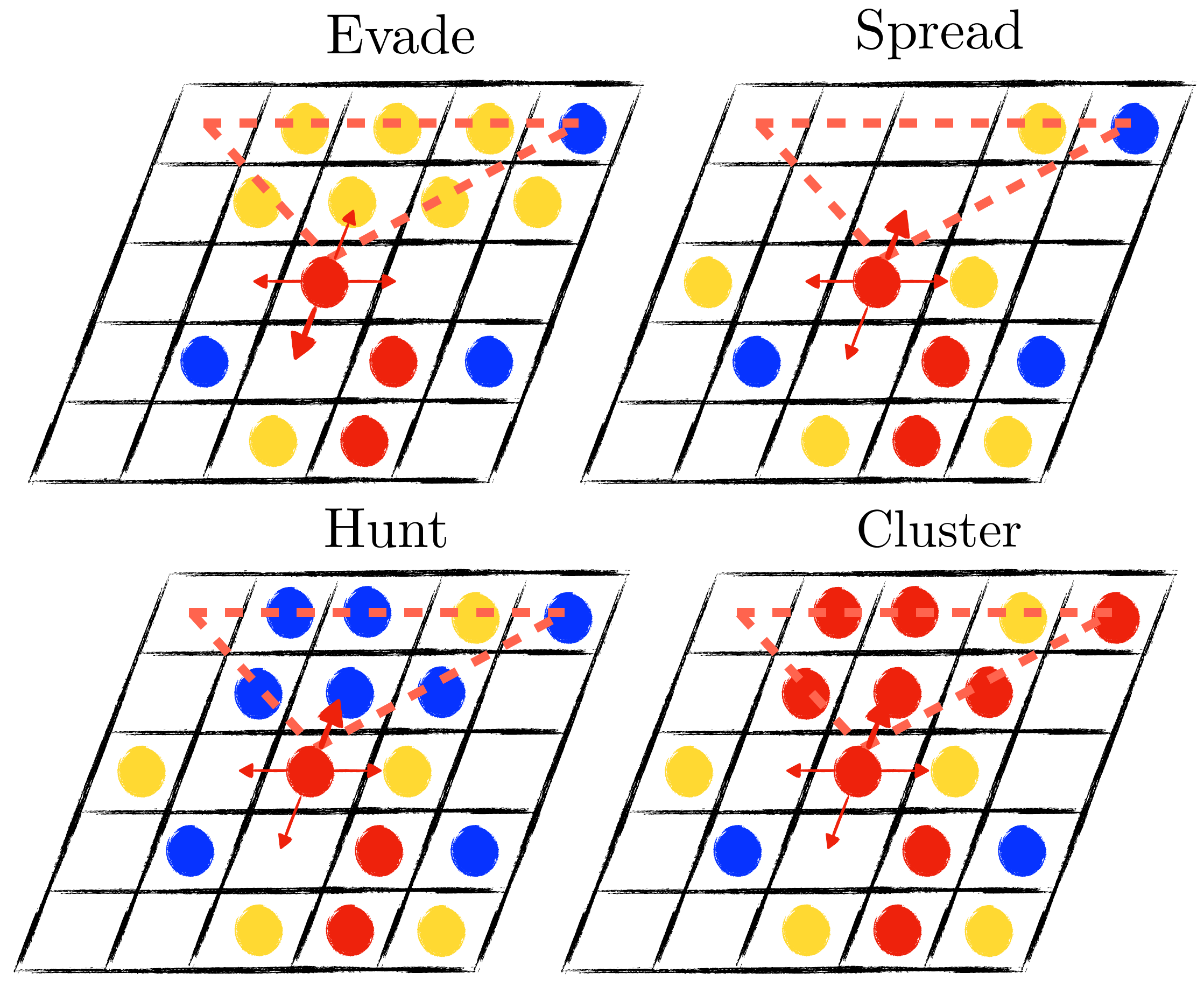}
    \caption{Illustration of strategies with directional biases, where
    a (smart) particle chooses to perform its action based on the local environment.
    These strategies are abbreviated as E (evade), S (spread), H (hunt) and C (cluster).
    If there is no directional bias then the strategy is ``non-directional''.  If all species behave identically then the strategy is ``null'' or ``symmetric'' (under permutation).
    }
    \label{fig:spatial_adaptation}
\end{figure}

\subsection{Directional movement strategies}\label{sec:loca_para}

As discussed in Sec.~\ref{sec:intro}, animals perceive their immediate surroundings and adjust their behaviours accordingly.  To incorporate this behaviour in our model, we allow the smart species $A$  to adjust the probabilities $P({\cal D})$ for the directions along which they perform actions.  
These probabilities will depend on the prey and predator individuals in its neighbourhood, as well as the empty spaces nearby, and on particles of the same species (which we call ``peer'' particles).  For species $A$, the prey is species $B$ and the predators are species $C$, recall \eqref{eq:actions-simple}.
Dynamics where particles choose their movement rates based on the local environment have been studied before, see for example~\cite{avelino2018directional, moura2021behavioural, tenorio2022adaptive, menezes2022local, menezes2022adaptive}.

We consider three types of behaviour for moving particles.  In the simplest case, we choose one of the four available directions at random: this is $P({\cal D})=P_0({\cal D})$ with
\begin{equation}
\label{eq:P0D}
P_0({\cal D}) = (1/4)
\end{equation}
For pure directional strategies (see below), each particle has a preferred direction ${\cal D}^*$ based on its environment (see below). Then we take $P({\cal D})=P_1({\cal D})$ with
\begin{equation}
\label{eq:P1D}
P_1({\cal D}) = 
\begin{cases} 
(1/4)+(\phi/4) , & \quad {\cal D}={\cal D}^* , \\ 
(1/4)-(\phi/12) , & \quad {\cal D}\neq {\cal D}^* \; .
\end{cases}
\end{equation}
with $0\leq\phi\leq3$ so that $\phi$ is the strength of the directional preference (it is possible to work with $-1\leq\phi\leq3$ but we restrict to positive $\phi$ so that ${\cal D}^*$ is indeed the preferred direction).

Finally, we consider mixed directional strategies in which particles have two preferred directions ${\cal D}^*_{1},{\cal D}^*_{2}$ with preferences $\phi_1,\phi_2$.  Then $P({\cal D})=P_2({\cal D})$ with
\begin{equation}
\label{eq:P2D}
P_2({\cal D}) = \frac14 + \frac{\phi_1}{4} \left( \delta_{{\cal D},{\cal D}^*_1} - \frac13 \right) +  \frac{\phi_2}{4} \left( \delta_{{\cal D},{\cal D}^*_2} - \frac13 \right)
\end{equation}
where $\delta_{{\cal D},{\cal D}^*}=1$ if ${\cal D}={\cal D}^*$ and zero otherwise, so that $P_2$ reduces to $P_1$ if $\phi_2=0$.  For mixed strategies we require $\phi_1+\phi_2\leq3$ and $\phi_1,\phi_2\geq0$.

\begin{figure*}[t!]
    \centering 
    \includegraphics[width=0.95\textwidth]{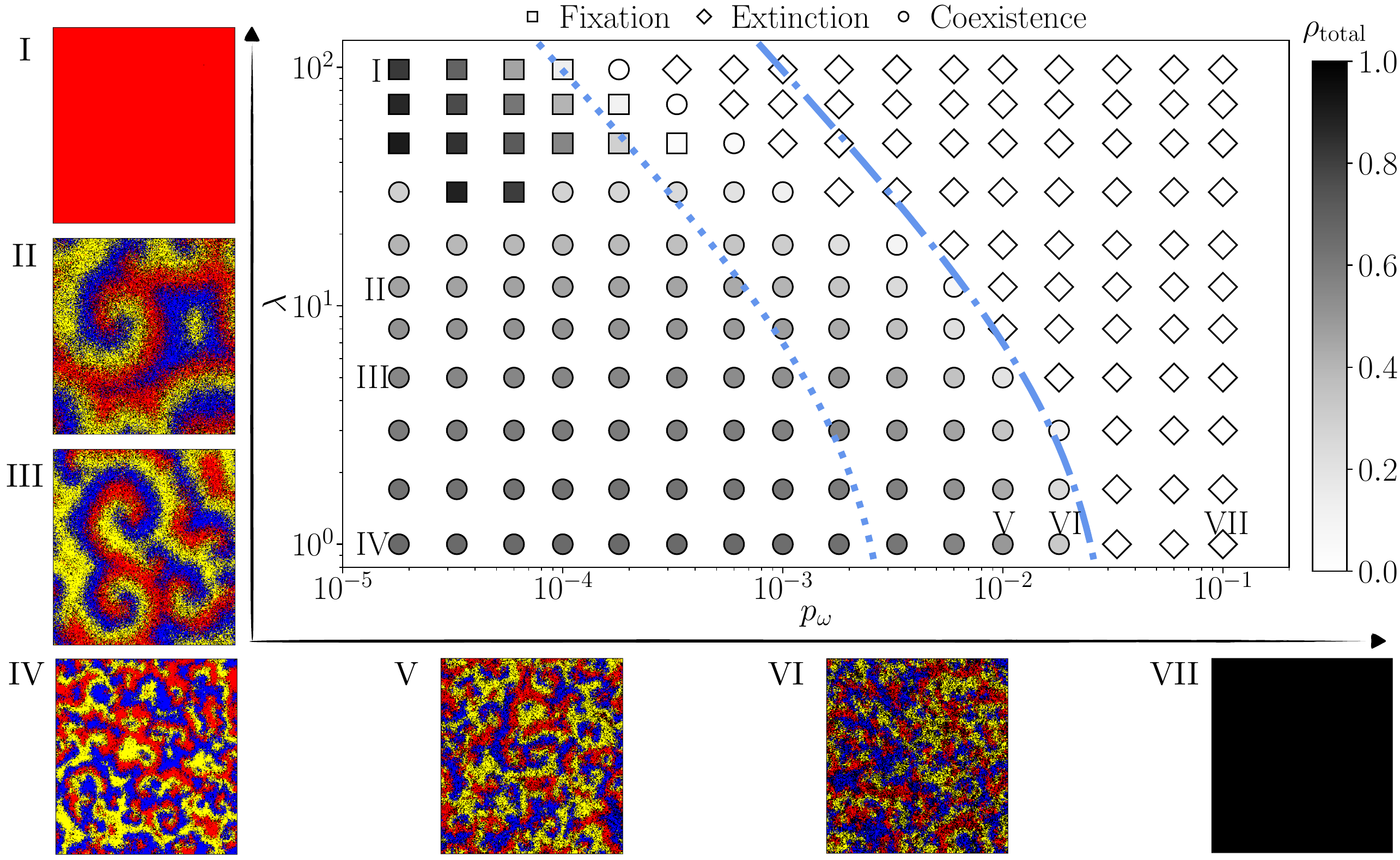}
    \caption{
    Phase behaviour of the system, as a function of $\lambda$ and $p_{\omega}$.  The shading shows the total particle density at each point in the control parameter space.  Snapshots I-IV show the coexistence-fixation transition. Snapshots IV-VII show the coexistence-extinction transition.  System size is $L=300$, results are based on simulations of $T = 10^5$ MCS.  Blue dotted and dash-dotted lines are curves of constant $\omega=0.01$ and $\omega=0.1$ respectively, see the text for a discussion (Sec.~\ref{sec:coexistence-fixation}).}
    \label{fig:phase_diagram}
\end{figure*}

To assign the preferred direction(s) for a particle at position $x$, we define its perception area to be a square of side $2{\cal R}+1$, centred at $x$.  See Fig.~\ref{fig:spatial_adaptation}, which also shows how this square is divided into four triangles, one associated with each direction ${\cal D}$.  An example pure directional strategy is  \emph{hunting} (H), where the preferred direction is assigned by counting the number of prey within each triangle and taking ${\cal D}^*$ to be the direction whose triangle has the maximal number of prey.  (In case of a tie, we take ${\cal D}^*$ to be one of the maximising directions, chosen uniformly at random.  Note also that the triangles overlap along the diagonals of the lattice: particles on those sites are counted in both triangles.)
We also define three other pure strategies: \emph{clustering} (C) where the preferred direction has the maximal number of peers; \emph{evasion of predators} (E), where the preferred direction has the minimal number of predators; and \emph{spreading} (S), where the preferred direction has the maximal number of empty site).
We define mixed strategies by combining two pure ones.  For example, the ``evasion \& hunting'' (E\&H) strategy assigns ${\cal D}^*_1$ according to the evasion strategy and ${\cal D}^*_2$ according to the hunting strategy.

The effectiveness of these strategies depends weakly on the perception range $\mathcal{R}$ (see App.~\ref{sec:perception_range}) so we fix the perception range $\mathcal{R}=3$ throughout the main text.
We emphasize again that for this work, the only species to have environmentally-dependent probabilities $P({\cal D})$ is the smart species $A$; the other species $B,C$ always choose their directions uniformly at random, as described in Sec.~\ref{sec:mc}.

\section{Phase Behaviour of the Model (symmetric case)}\label{sec:phase_behaviour}
In this section, we describe the phase behaviour of the model for parameters where all three species have the same dynamics.  We refer to this as the \emph{symmetric case} because the behaviour is invariant under cyclic permutation of the species.
Specificall,y we take
\begin{align}
\sigma_A & = \sigma_B = \sigma_C = 1
\nonumber\\
\mu_A & = \mu_B = \mu_C = 1
\nonumber\\
\lambda_A & = \lambda_B = \lambda_C = \lambda, 
\end{align}
and there is no directional preference ($\phi=0$ or equivalently $P({\cal D})=1/4$ always).
The time step is $\tau = \frac{1}{\lambda+3}$ which ensures that all probabilities in \eqref{eq:actions_prob} are between $0$ and $1$.
We vary the selection rate $\lambda$ and the spontaneous death probability $p_\omega=\tau\omega$.
The results demonstrate the differences between the model of this work and the (original) RPS model of~\cite{reichenbach2007mobility, reichenbach2008self}.  They also serve as a baseline for later Sections where the symmetry among the 3 species is broken.  

\subsection{Phase Diagram}\label{sec:phase_transition}

We ran simulations of $10^5$ MCS for systems of of size $L\times L=300\times 300$, and a range of parameters $(p_\omega,\lambda)$.  We allow $9\times 10^4$ MCS for the system to settle into its steady state, after which we recorded particle densities $(\rho_A, \rho_B, \rho_C)$ of the system, which is averaged over the time period $9\times 10^{4} < t <  10^5$ MCS.
Fig.~\ref{fig:phase_diagram} shows results, including a phase diagram, and snapshots of the system's final configuration at the selected state points.
Note that if a species dies out (no remaining individuals) then no new particles of that species can be born, so the number of species in the system can never increase.  

The resulting phase diagram features three distinct phases which are called fixation (only one species is present at the final time), coexistence (three species are present) and extinction (all sites are empty).   As in \cite{reichenbach2007mobility}, it is not possible that two species survive at long times since one of them will always dominate the other, which leads to either fixation or extinction.  The total density $\rho_{\rm total}$ is also indicated: this is zero in the extinction phase.

The fixation phase occurs for large movement rate $\lambda$ and small death probability $p_\omega$.  
Since the system is symmetric, the species that survives in this phase is completely random.  Reducing the movement rate $\lambda$ favours the coexistence phase, in which spiral patterns appear, characterised by a length scale that grows with $\lambda$.  This is the same behaviour observed in the RPS model of~\cite{reichenbach2007mobility,reichenbach2008self, reichenbach2008instability}, consistent with the fact that our model reduces to theirs on setting $p_\omega=0$.  (In that case, the fixation phase also has $\rho_X=1$ for the surviving species.)

On increasing $p_\omega$, the behaviour changes qualitatively because the death process favours the extinct state.  Indeed for $p_\omega\gtrsim 10^{-3}$, one still has the coexistence phase for small $\lambda$, but increasing $\lambda$ leads to extinction instead of fixation.  As noted above, the reason is that if only a single species survives (fixation phase) then all particles will end up hungry, reducing their reproduction rate.  Then the whole population tends to collapse into the extinct state.  This illustrates how the combination of the death process and the hunger levels leads to a more complex ecosystem, where different species rely on each other for survival.  Increasing the death rate also tends to disrupt the spiral patterns, compare snapshots IV, V, and VI in Fig.~\ref{fig:phase_diagram}.  Eventually, the system fragments into irregular clusters of each species~\cite{bhattacharyya2020mortality, islam2022effect, gopaoco2023role}.

\subsection{Transitions between coexistence and fixation/extinction phases}\label{sec:coexistence-fixation}
The transition between coexistence and fixation phases has been the focus of many previous studies~\cite{reichenbach2007mobility, reichenbach2007noise, reichenbach2008self, reichenbach2008instability, knebel2013coexistence}.  The length scale of the spiral patterns grows with $\lambda$ until it becomes system-spanning (see snapshots IV, III, II, and I in Fig.~\ref{fig:phase_diagram}).  Moreover, these spirals are associated with oscillations in species' populations, and for system-spanning spirals, these oscillations are large enough that one species may die out.  This leads to an explosion in the population of its prey species, which then wipes out the remaining species (its prey).  This leads to fixation.  

Note however that since the number of species can never increase, the coexistence phase is necessarily ``metastable'': for fixed system size and with sufficiently long simulation, the system will eventually end up in the fixation phase~\cite{reichenbach2007mobility, reichenbach2007noise, reichenbach2008self, reichenbach2008instability}.  Nevertheless, the transition between coexistence and fixation is well-defined in the limit of large system size, where it can be characterised via the scaling with $L$ of the time to reach fixation~\cite{reichenbach2007noise}.  However, the inherent metastability of the coexistence phase must be borne in mind when analysing simulation behaviour, this will become clear in later Sections.

In contrast to the coexistence-fixation transition, the transition to an extinct state is not present in the RPS model of~\cite{reichenbach2007mobility, reichenbach2007noise, reichenbach2008instability, reichenbach2008self, berr2009zero}.  (This transition relies on the death process and the hunger levels.)  Unsurprisingly, increasing the death probability $p_\omega$ tends to reduce the total population: this eventually collapses because dilute systems make it increasingly hard for particles to find prey, leading to hunger, reduced reproduction and hence extinction. 
Another interesting effect of increased $p_\omega$ is the loss of coherence in the spiral pattern (panels V and VI in Fig.~\ref{fig:phase_diagram}).  

{The same transition (from coexistence to extinction) also appears on increasing mobility $\lambda$ at fixed $p_\omega$. 
As in for the coexistence-fixation transition, it is also important that increased mobility leads to longer-ranged spatial correlations and large fluctuations, so that species are more likely to die out via random fluctuations.
To understand the shape of the phase boundary, we recall that the time step $\tau$ depends on $\lambda$ in the results of Fig.~\ref{fig:phase_diagram}, so fixing $p_\omega$ does not correspond to a fixed rate $\omega$, due to \eqref{eq:om-p-om}.  Lines of fixed $\omega$ are shown in blue in Fig.~\ref{fig:phase_diagram}, these indicate that transition from coexistence to extinction takes place at a death rate $\omega=\omega_{\rm x}$ that is between $0.01$ and $0.1$, depending weakly on $\lambda$.  This indicates that the most important control parameters of the model are ratios of rates, for example $\omega/\mu$ sets the balance between reproduction and spontaneous death (note that $\mu=1$ is constant in Fig.~\ref{fig:phase_diagram}).  In later Sections we keep a fixed time step $\tau$ so it is equivalent to fix either $\omega$ or $p_\omega$.}

\section{Learning by a ``smart'' species}\label{sec:learning}

\subsection{Motivation}

The central question of this work is how a privileged (smart) species can adjust its behaviour, in order to maximise its population.  (Specifically, we adjust the parameters $\lambda_A,\sigma_A,\phi$ as well as adopting different strategies when choosing the preferred direction ${\cal D}^*$.)  In principle, this question could be addressed in simulation by scanning the various parameters.  Instead, we adopt a different approach based on reinforcement learning (RL).  The method is detailed below: as usual in RL, the main idea is that we mostly run simulations at parameters that have previously been found to be good, but this is supplemented by exploratory searching, to find other regions of parameter space that might be even better.

A priori, this method seems promising for two reasons: Firstly, we expect it to be more efficient than parameter scanning, in the context of our simulation study.  Secondly, it may mimic the mechanisms by which organisms actually learn and evolve, in the context of real ecosystems~\cite{pearce2013animal, kao2014collective, sasaki2017cumulative}.  Note however that the method we employ here does not involve learning by individual particles: the value function is defined at the level of the species, and it is assumed that individuals act according to some shared processing of this information.  Such ideas have provided valuable insight into many social behaviours of animals such as ants and bees~\cite{wilson1989reviving, seeley1989honey, parrish1999complexity, seeley2009wisdom} and it sometimes termed ``social learning''~\cite{galef2005social, reed2010social, whiten2017social}.

\begin{figure}[t!]
    \centering 
    \includegraphics[width=0.47\textwidth]{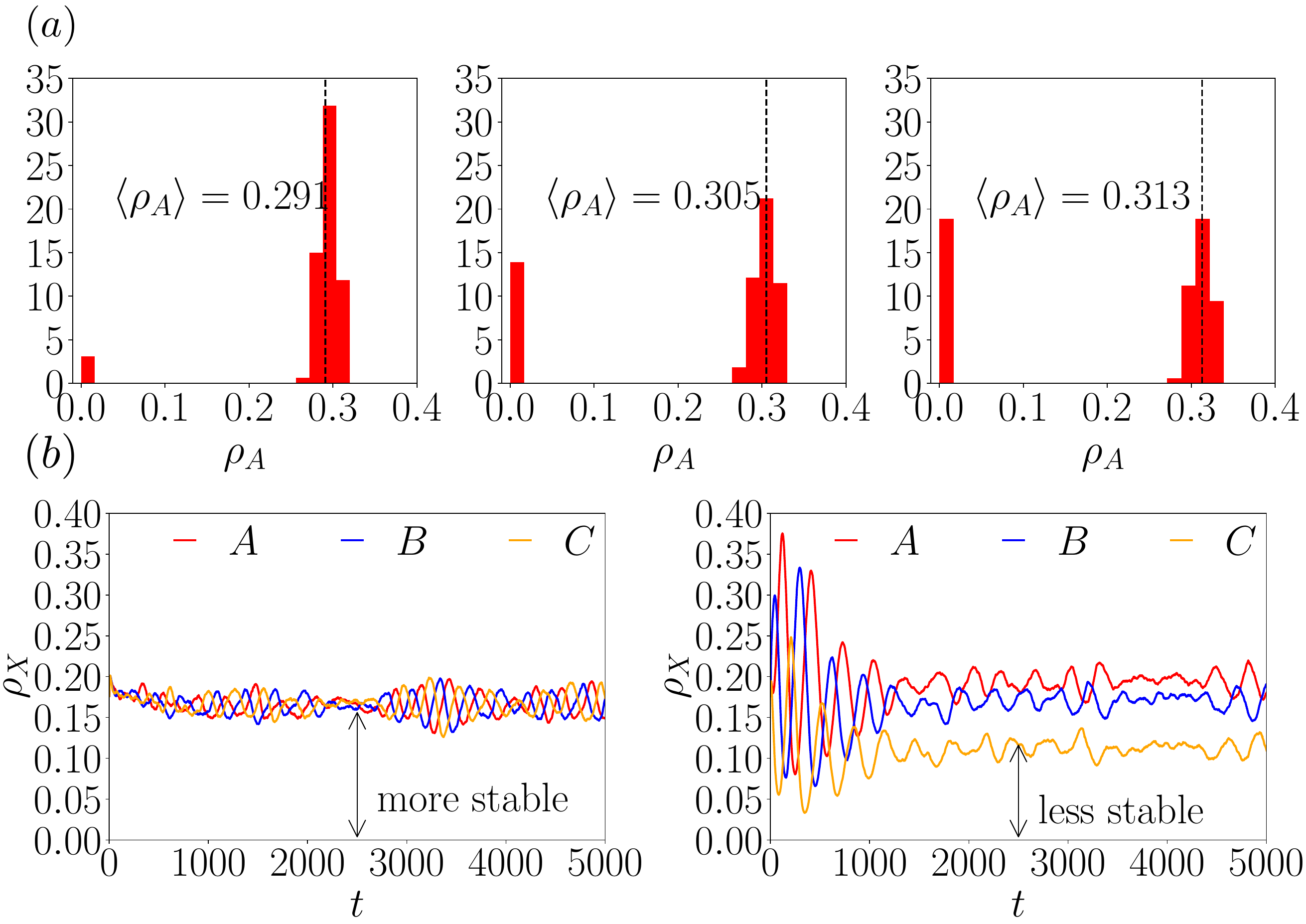}
    \caption{(a) Probability distribution of $\rho_A$ for $\sigma_A=0.31, 0.25, 0.22$ (from left to right) and $\lambda_A=1.8$. Also, $p_{\omega}=0.01$, and $T=2\times10^5$ MCS, other parameters are fixed as in Table~\ref{tab:fixed_para}.  Each distribution is obtained from $100$ simulations.
    (b) Time series for $\sigma_A=1.0$ (symmetric) and $\sigma_A= 0.5$ (asymmetric) with $\lambda_A=1.0$ and $p_{\omega}=0.015$.  In the latter case, $\rho_A$ is increased at the cost of $\rho_C$.
    }
    \label{fig:prob_setup}
\end{figure}

\begin{table}[]
\begin{tabular}{|l|l|}
\hline
\multicolumn{1}{|c|}{Fixed Parameters}                               & \multicolumn{1}{c|}{Value} \\ \hline
Time step, $\tau$              & $(2/9)$                        \\ \hline
Reproduction rates, $\mu_A, \mu_B, \mu_C$ & $1$                        \\ \hline
Predation rates, $\sigma_B, \sigma_C$                    & $1$                        \\ \hline
Movement rates, $\lambda_B, \lambda_C$ & $1$                        \\ \hline
Hunger progression probability, $p_h$                                & $0.02$                     \\ \hline
Hunger reproduction factor, $\mathcal{H}$                            & $0.02$                     \\ \hline
System size, $L$                                &  $120$$^*$                     \\ \hline
Death probability, $p_{\omega}$ & $0.005$ (sparse) \\ & $0.015$ (crowded) \\ \hline
\end{tabular}
\caption{The list of fixed parameters in the RL calculations, and their values.
\\ $^*$ We use $L=120$ for optimization calculations using RL.  The results of Secs.~\ref{sec:phase_behaviour} and~\ref{sec:survival_of_weakest} used larger lattices, $L=300$.
}
\label{tab:fixed_para}
\end{table}

\subsection{Optimisation problem}

We use RL to optimise the population of the smart species ($A$). 
As noted above, we choose the death probability and the hunger parameters such that fixation is not possible, so this optimum is achieved in the coexistence phase.  
However, we also explained in Sec.~\ref{sec:phase_transition} that the coexistence phase is necessarily ``metastable'', and finite systems must always enter the extinct state at some sufficiently long time.  To illustrate this, Fig.~\ref{fig:prob_setup}(a) shows histograms of $\rho_A$ obtained after simulation of $T=2\times 10^5$ MCS.  The distribution has two peaks: one at $\rho_A=0$ corresponding to extinction, and one at $\rho_A>0$, corresponding to coexistence.  Since extinct systems never recover, increasing $T$ always increases the probability of extinction.  

To avoid problems associated with this effect, we set up our optimisation problem as follows.  Define
\begin{equation}
r(t) = 
\begin{cases} 
\rho_A(t) , &  \quad \rho_A(t) \geq \delta \; , \\ 
-1 , & \quad \rho_A(t) < \delta \; .
\end{cases}
\label{eq:rt}
\end{equation}
such that $r(t)$ is the $A$-population for systems in the coexistence phase, but $r(t)=-1$ if species $A$ dies out, or if its population is lower than a threshold $\delta$.   We take $\delta=0.05$, the idea is that for these small populations the species is likely to be on the pathway to extinction, even if this state has not been reached.

\begin{figure}[t!]
    \centering 
    \includegraphics[width=0.4\textwidth]{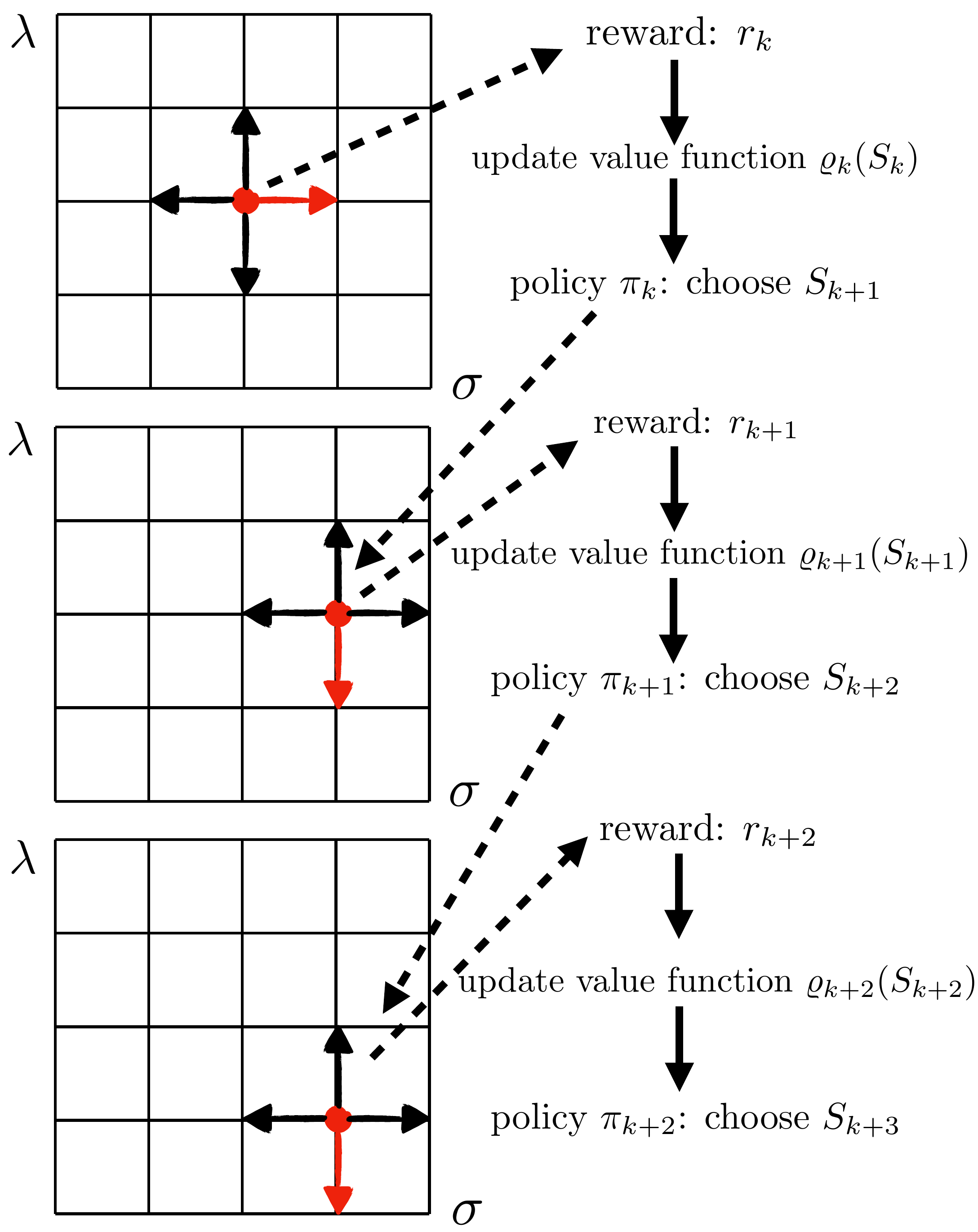}
    \caption{
    Illustration of the RL algorithm.  At each step $k$, the algorithm receives a reward $r_k$ and updates its state to $S_{k+1}$ using the $\epsilon$-greedy policy \eqref{eq:greedy}.  
    The red arrows indicate the chosen transition in each step.
    }
    \label{fig:mechanism}
\end{figure}

We write $\langle O(t) \rangle_{\rm coex}$ be the average of an observable $O(t)$, for a system started in the coexistence phase at time $t=0$.  
We aim to optimise a set of parameters $\sigma_A,\lambda_A,\dots$ and we write
$S=(\sigma_A,\lambda_A,\dots)$ for a particular choice of of these parameters.  Then the value function for our optimisation is
\begin{equation}
\varrho(S) = \langle r(T) \rangle_{\rm coex}
\label{eq:value}
\end{equation}
where the parameter $T$ is taken small enough that this $\varrho$ has a positive maximum (corresponding to a metastable coexistence phase) but large enough to allow exploration of this phase.  Compared to taking simply $r(t)=\rho_A(t)$, the definition \eqref{eq:rt} penalises parameters $S$ where the system has a significant probability of extinction within the time $T$.  (One might also consider larger penalties, by replacing the value $-1$ in \eqref{eq:rt} by $-r_{\rm pen}$ with some $r_{\rm pen}>1$.  We expect similar results in this case.)

The optimisation is performed over the movement and predation rates of species $A$.  Other parameters are fixed, with values given in Table~\ref{tab:fixed_para}.  These values are not fine-tuned and we expect the behaviour observed here to be robust across a range of parameters, within the physical constraints already discussed.

To better understand the optimisation problem, we show example time series for the populations in Fig.~\ref{fig:prob_setup}(b), for a system in the coexistence state. The populations show the characteristic oscillations associated with predator-prey dynamics~\cite{hofbauer1998evolutionary,dobramysl2018stochastic}.
The competitive nature of the dynamics and the volume exclusion constraint both mean that a large population for one species tends to occur at the expense of the others~\cite{kerr2002local, berr2009zero, avelino2018directional, menezes2019uneven}.  In particular, Fig.~\ref{fig:prob_setup}(b) shows that reducing the predation rate from $\sigma_A=1$ to $\sigma_A=0.5$ means that species $A$ gains larger population at the expense of species $C$.  This higher $\rho_A$ is the result of the ``survival of the weakest'' which we will discuss in Sec.~\ref{sec:survival_of_weakest} below.   At this point, we note that that if species $A$ achieves a large population, its prey species $B$ is likely to be less numerous.  However, if $\rho_B$ is small, this species runs the risk of dying out, which leads in turn to the collapse of the whole ecosystem [extinction of all species and $r(t)=-1$].  
Hence, the optimisation problem for species $A$ is twofold: how can it learn an advantageous survival strategy that improves its population density, while still sustaining a stable prey population, and maintaining the system's biodiversity?  In other words, the ``smart'' species need to balance optimising its population and keeping ecosystem sustainable.

\begin{figure*}
\begin{minipage}{\linewidth}
\begin{algorithm}[H]
  \caption{Reinforcement learning of $\varrho(S)$}
  \label{alg:bandit}
   \begin{algorithmic}[1]
   \State {initialise} model parameters.
   \State {initialise} $\hat\varrho(S)=-1$ and $n(S)=1$ for all $S$.
   \For{$e=1...E$}
   \Repeat
   \State initialise the system at random state point $S_1$ and simulate time $T$.
   \Until a state point is found such that $\rho_A(T)>0$.
   \For{$k=1...K$}
   \State reset the hunger level of all the particles in the system to 0.
   \State simulate the system at state $S_k$ based on Algorithm~\ref{alg:MC}.
   \State measure reward $r_k$.
   \State update value estimate $\hat\varrho(S_k)$ using Eq.~\eqref{eq:value-upd}.
   \State update $n(S_k)\leftarrow n(S_k)+1$.
   \If{ $r_k>0$ (species $A$ has not died out)}
   \State choose new state point $S_{k+1}$ based on policy in Eq.~\eqref{eq:greedy}.
   \Else
   \State terminate current episode.
   \EndIf
   \EndFor
   \EndFor
   \end{algorithmic}
\end{algorithm}
\end{minipage}
\end{figure*}

\subsection{Learning Algorithm}

We optimise the value function $\varrho$ over the parameters $S$ by a type of multi-armed bandit algorithm~\cite{berry1985bandit, auer2002nonstochastic, mahajan2008multi, kuleshov2014algorithms, sutton2018reinforcement}.  However, we insist (contrary to standard algorithms) that all updates to $S$ are small, in order to mimic the predominantly incremental process of collective learning and adaptation in evolutionary biology~\cite{mccloskey1989catastrophic, kudithipudi2022biological, van2022three}, see also~\cite{sumpter2010collective, pearce2013animal, kao2014collective, staddon2016adaptive, sasaki2017cumulative}.

We optimise over a set of three or four parameters, whose values are discretised on a grid.  Each point on the grid is  a \emph{state point} $S$, recall \eqref{eq:value}.  Our aim is to learn the value function $\varrho(S)$ in the vicinity of the optimal state point $S^*$.  The RL method achieves this via a function $\hat\varrho(S)$ which is an estimator for $\varrho(S)$.    

The method is illustated in Fig.~\ref{fig:mechanism}, it proceeds in steps indexed by $k=1,\dots,K$, which are further organised into training episodes, indexed by $e=1,\dots,E$.  On step $k$ the state point is $S_k$, and a simulation is run at this state point in order to improve $\varrho(S_k)$.  A new state point $S_{k+1}$ is chosen on the basis of the estimated values, and the method continues.  The constraint of incremental updates to $S$ means that $S_{k+1}$ is always a neighbour of $S_k$ on the parameter grid.  In addition to the current estimate of $\varrho(S)$, we define variables $n(S)$ to keep track of the number of simulations that have been performed at state point $S$ (this is relevant for the uncertainty of the estimate $\hat\varrho(S)$).

This scheme is formalised in Algorithm~\ref{alg:bandit}, and we now describe this method.  The reward estimates are initialised to the arbitrary value $\hat\varrho(S)=-1$ for all $S$, and all $n(S)$ are initialised to $1$.   Each training episode begins with a random state point $S_1$ that supports a finite population of $A$.  (This is achieved by choosing a random state point $S_{\rm init}$ and simulating $T$ MCS: if the final population is non-zero then take $S_1=S_{\rm init}$, else choose another state point $S_{\rm init}$, and repeat this procedure until a finite population is found.) Each episode includes many steps of the algorithm, and every step involves a simulation of $T$ MCS.  The initial condition of each simulation is taken as the final condition of the last one, so one may think of a species adjusting its behaviour in order to find effective strategies.  However, if the population of species $A$ drops below $\delta$ at any point then the episode ends and the next episode starts with a new random state point $S_1$.

During step $k$ the parameters are $S_k$.  The step consists of $T$ MCS and we average the reward $r(t)$ in \eqref{eq:rt} over the final $T_{\rm meas}$ MCS, and denote its value by $r_k$.  Then we update our estimate of the relevant value function as
\begin{equation}
\hat\varrho(S_k) \leftarrow \hat\varrho(S_k) + \frac{1}{n(S_k)} [ r_k - \hat\varrho(S_k) ]
\label{eq:value-upd}
\end{equation}
and we also update $n(S_k) \leftarrow n(S_k)+1$.   This update ensures that
\begin{equation}
\hat\varrho(S) = \frac{1}{n(S)} \sum_{i=1}^{n(S)} r(S,i)
\end{equation}
where $r(S,i)$ is the reward for the $i$th simulation at state point $S$.  {(This sum generically includes contributions from all episodes, note however that $r(S,1)=-1$ is fixed by initialisation and does not correspond to an actual simulation.  Results depend weakly on this choice.)}
The more simulations are performed at state point $S$, the more accurately $\hat\varrho(S)$ approximates the value function $\varrho(S)$, which is the average reward.
The above-described process is called value evaluation.

It remains to describe the method of choosing $S_{k+1}$, which is called the learning policy.  
As noted above, the only possible choices for $S_{k+1}$ are adjacent to $S_k$ on the parameter grid. 
(We do not allow $S_{k+1}=S_k$.) We write ${\cal  N}_k$ for the set of possible choices and we
take the $\epsilon$-greedy policy
\begin{equation}
    S_{k+1}= \begin{cases}
     \argmax_{S\in {\cal N}_k} \hat\varrho(S) &  \text{with prob.}\, 1-\epsilon \\
     \text{random element of}\, {\cal N}_k & \text{with prob.}\, \epsilon \\
\end{cases}, 
\label{eq:greedy}
\end{equation}

This procedure can be described in the framework of Markov decision processes~\cite{littman1994markov, kaelbling1996reinforcement, sutton2018reinforcement}.  Within each episode, we consider a trajectory as a sequence of state points, and associated value estimates $S_1,r_1,S_2,r_2,\dots$.  In the context of Markov decision processes, the state-action on the $k$th step simply reduces to the state point $S_k$ (the standard multi-armed bandit has a similar feature).

The separation of the training process into episodes aids exploration of the state space by resetting to a completely random state point at the start of each episode, as well as providing a mechanism for the system to recover from extinction events.  A side-benefit is that it aids the analysis of convergence of the learning process, see below.

\begin{table}[]
\begin{tabular}{|l|l|}
\hline
\multicolumn{1}{|c|}{Parameters}                                                                          & \multicolumn{1}{c|}{Value} \\ \hline
Number of episodes, $E$                                                                                    & $2000$                            \\ \hline
Number of steps, $K$                                                                                      & $20$ or $30$                     \\ \hline
Greedy factor, $\epsilon$                                                                                 & $0.2$                            \\ \hline
Tolerance in reward calculation, $\delta$ \hspace{18pt}                                                                                      & $0.05$                     \\ \hline
Simulation time (MCS), $T$                                                                                      & $5000$                     \\ \hline
Measurement period, $T_{\rm meas}$                                                                         & $2000$                     \\ \hline
\end{tabular}
\caption{The parameters used in the learning algorithm and their values.}
\label{tab:learn_para}
\end{table}

\subsection{Algorithm implementation and convergence}

Having described the general algorithm, we now discuss its application in practice.  We keep most parameters fixed while optimising relevant parameters for species $A$.  The fixed parameters are summarised in Tab.~\ref{tab:fixed_para}.  [The time step is now fixed at $\tau=(2/9)$ which allows $\lambda_A$ to be adjusted at fixed $\tau$, recall the probabilities in \eqref{eq:actions_prob} must all be positive.]

For illustration, consider a pure directional strategy for species $A$ as described in Sec.~\ref{sec:loca_para}, for example the hunting strategy.  We aim to optimise three parameters $\sigma_A,\lambda_A,\phi$.  The grid for the parameters is defined as follows:
$\sigma_A$ and $\lambda_A$ are varied between $0$ to $2$ with grid spacing $0.2$, but we restrict $\sigma_A + \lambda_A \leq 2.5$ for numerical convenience.  (This reduction of the search space does not affect the optimal strategy.) The directional parameter $\phi$ runs from $0$ to $3$ with grid spacing $0.25$.
For mixed strategies as in Eq.~\ref{eq:P2D} we optimise four parameters $\sigma_A,\lambda_A,\phi_1,\phi_2$ where the grid spacing for $\phi_1,\phi_2$ is again $0.25$, we restrict $\phi_1,\phi_2>0$ and $\phi_1+\phi_2\leq 3$.

The main parameters of the RL algorithm are given in Table~\ref{tab:learn_para}.  {The number of episodes $E$ is chosen to be $2000$ to ensure convergence of the value function.  The number of steps $K=20$ for pure directional strategies and $K=30$ for mixed directional strategies, this ensures that in each episode the algorithm sufficiently explores the grid of state points, given that the dimensionality of this grid is larger for the mixed strategies.  The greedy factor $\epsilon=0.2$ ensures the balance between exploration vs. reinforcement.  Simulation time $T$ and measurement period $T_{\text{meas}}$ are chosen to ensure reward is obtained in a steady state.}

\begin{figure}[t!]
    \centering 
    \includegraphics[width=0.48\textwidth]{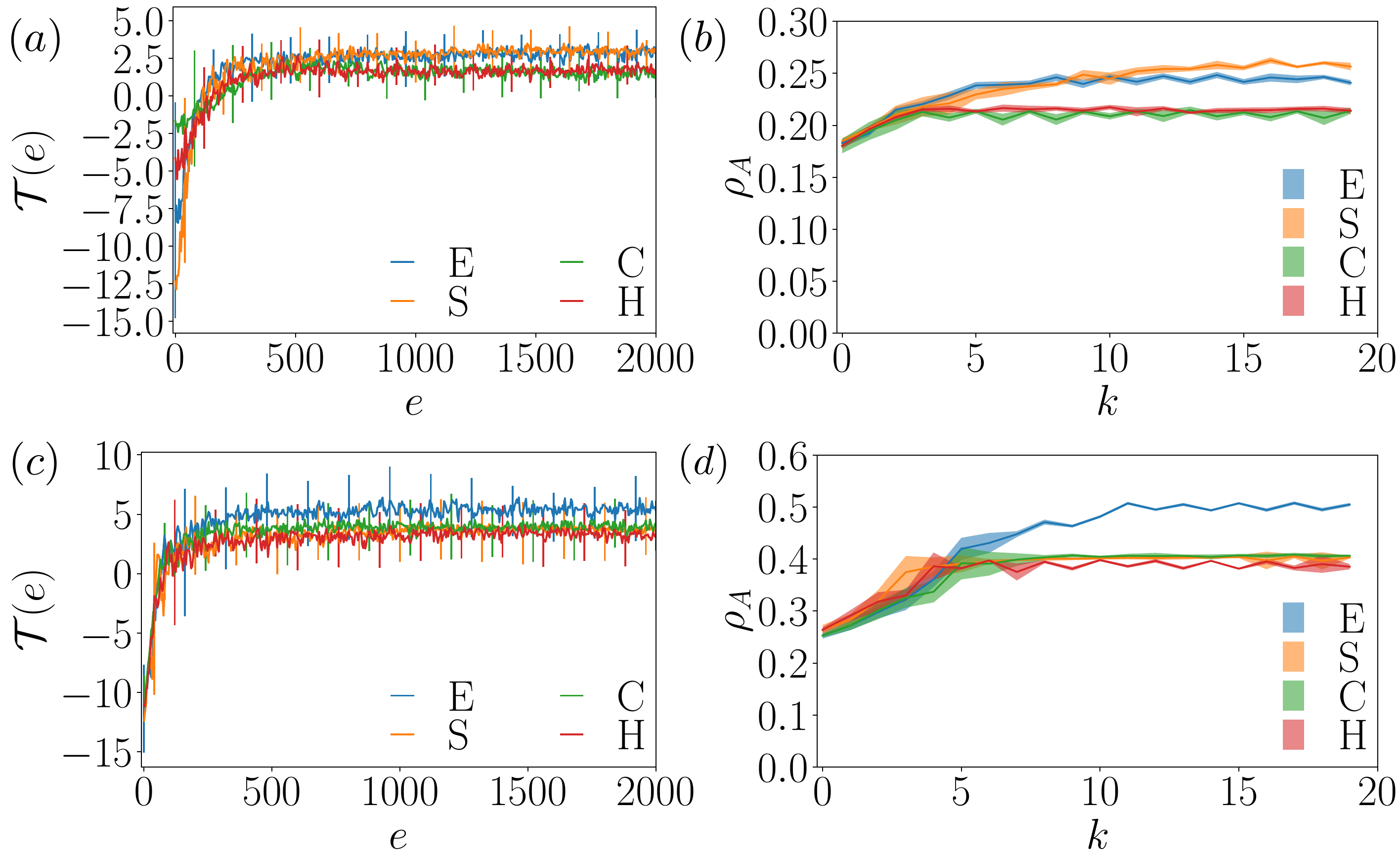}
    \caption{%
    (a)~The integrated reward $\mathcal{T}(e)$ obtained in each episode as a function of episodes for different (pure) strategies at $p_{\omega}=0.015$.  The data is shown as the average of $10$ runs.
    (b)~Evolution of $\rho_A$ for RL runs starting from a pre-learned value function at $(\sigma_A=1.0, \lambda_A=1.0, \phi=0)$ with $p_{\omega}=0.015$.
    (c,d)~Similar data to (a,b) at $p_{\omega}=0.005$.
    }
    \label{fig:learning_curve}
\end{figure}

As the algorithm runs, the value estimates $\hat\varrho$ converge to the value function $\varrho$, and the distribution of visited state points also converges to a steady state. 
To assess the convergence of our algorithm, we introduce the integrated reward for episode $e$:
\begin{equation}
\mathcal{T}(e)=\sum_{k=1}^K \hat\varrho(S_k,e,k)
\end{equation}
where $\hat\varrho(S,e,k)$ is the estimated value for state point $S$ after step $k$ of episode $e$.
If the episode ends due to extinction (before $K$ steps have been carried out) then we truncate the sum accordingly.
Note that ${\cal T}(e)$ depends on the value estimates, as well as the state points that are visited during the episode.
The value of ${\cal T}(e)$ fluctuates between episodes because the $S_k$ are stochastic, but there is no net drift.  See Fig.~\ref{fig:learning_curve}(a), which is obtained for the four pure directional strategies by running the whole algorithm 10 times and averaging the results for ${\cal T}(e)$. 

In addition Fig.~\ref{fig:learning_curve}(b) illustrates the operation of the $\epsilon$-greedy policy.  We take the learned $\hat\varrho$ from a previous run of the RL algorithm; then we initialise the system at $S=(\sigma_A,\lambda_A,\phi)=(1,1,0)$ and run a single episode, computing the population $\rho_A$ of species $A$ at the end of each step. This procedure is repeated averaged over 10 independent runs (always starting with the same pre-learned value estimates $\hat\varrho$).  The results show that the greedy policy successfully increases the population of the smart species, via parameter optimisation.
Figs.~\ref{fig:learning_curve}(c,d) demonstrate convergence and successful optimisation for a smaller value of $p_{\omega}$, demonstrating the robustness of the method.

\section{Results -- optimal strategies}
\label{sec:res-RL}

\subsection{Optimisation by RL}

The RL algorithm yields value estimates $\varrho$ from which we infer the (estimated) optimal state point
\begin{equation}
S^* = \argmax_{S} \hat\varrho(S) \; .
\end{equation}
In this Section, we explore the optimal state points that are obtained when optimising parameters for the various directional strategies introduced in Sec.~\ref{sec:loca_para}.
We consider two different death rates $p_{\omega}=0.015$ and $p_{\omega}=0.005$, to show the robustness of our method and investigate the environment dependence of adaptive strategies.  
The larger death rate $p_{\omega}=0.015$  leads to a lower total population density so we refer to this as the sparse case; the other value $p_{\omega}=0.005$ is the crowded case.

\begin{table*}[t]
\input{tab_opt.tex}
\caption{Optimal parameters for different strategies. }
\label{tab:optimal_para}
\end{table*}

\begin{figure}[t!]
    \centering 
    \includegraphics[width=0.49\textwidth]{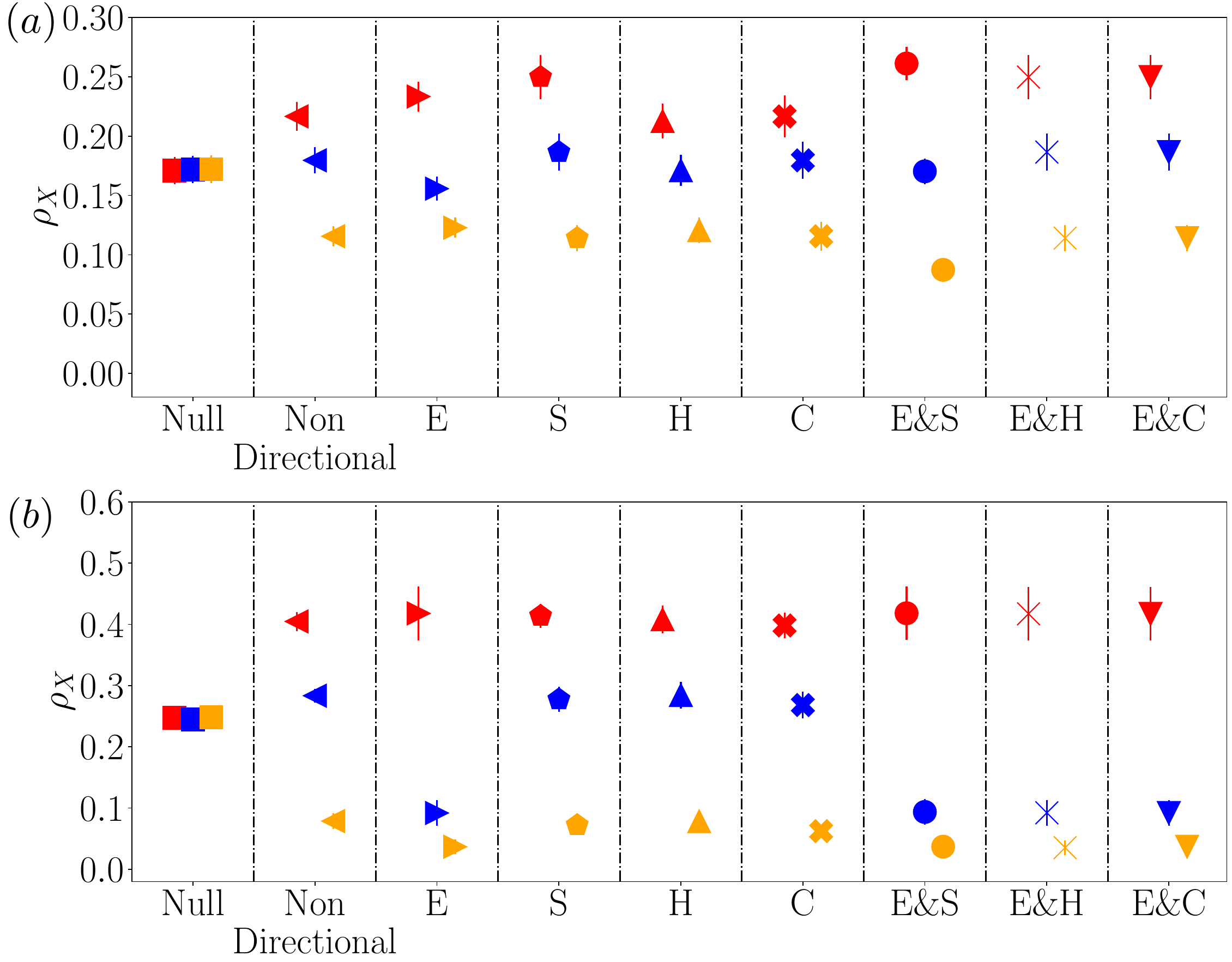}
    \caption{
    (a)~Optimised steady-state population densities for different survival strategies at $p_{\omega}=0.015$ averaged over 5 simulations, labelled according to Fig.~\ref{fig:spatial_adaptation}.  
    The E\&S strategy yields the highest $\rho_A$. 
    Error bars show the standard error of the mean.   
    (b)~Optimised steady state population densities for different survival strategies at $p_{\omega}=0.005$.  The evade strategy yields the highest $\rho_A$.
    }
    \label{fig:density_compare}
\end{figure}

As above, we fix a pure directional strategy for the $A$ particles and perform three-parameter optimisation for $S=(\sigma_A,\lambda_A,\phi)$.  We repeat this procedure for the four possible directional strategies (Sec.~\ref{sec:loca_para}) as well as for the non-directional strategy ($\phi=0$). 
For each strategy, we identify the corresponding $S^*$ and we perform MC simulations (without further learning) to estimate the species' populations $\langle \rho_X\rangle$ for $X=A,B,C$.  We also consider the symmetric (``null'') case in which species $A$ behaves identically to $B,C$, that is $(\sigma_A,\lambda_A,\phi)=(1,1,0)$.  

Results are shown in Fig.~\ref{fig:density_compare}, the densities obtained in each case are averaged over 5 simulations.  (All of these systems remained in the coexistence state throughout, there was no extinction or fixation.)   The learned (optimised) strategies generically lead to larger $\rho_A$ than the symmetric (null) case, as they should.  (The Figure also shows results for mixed strategies, these are discussed below.)  Among pure strategies, spreading leads to the largest $\rho_A$ in the sparse case ($p_{\omega}=0.015$).  For the crowded case, the picture is less clear-cut: the evasion strategy has the largest mean population but the other pure-directional strategies perform similarly well, as does the non-directional one.

\begin{figure}[t!]
    \centering 
    \includegraphics[width=0.48\textwidth]{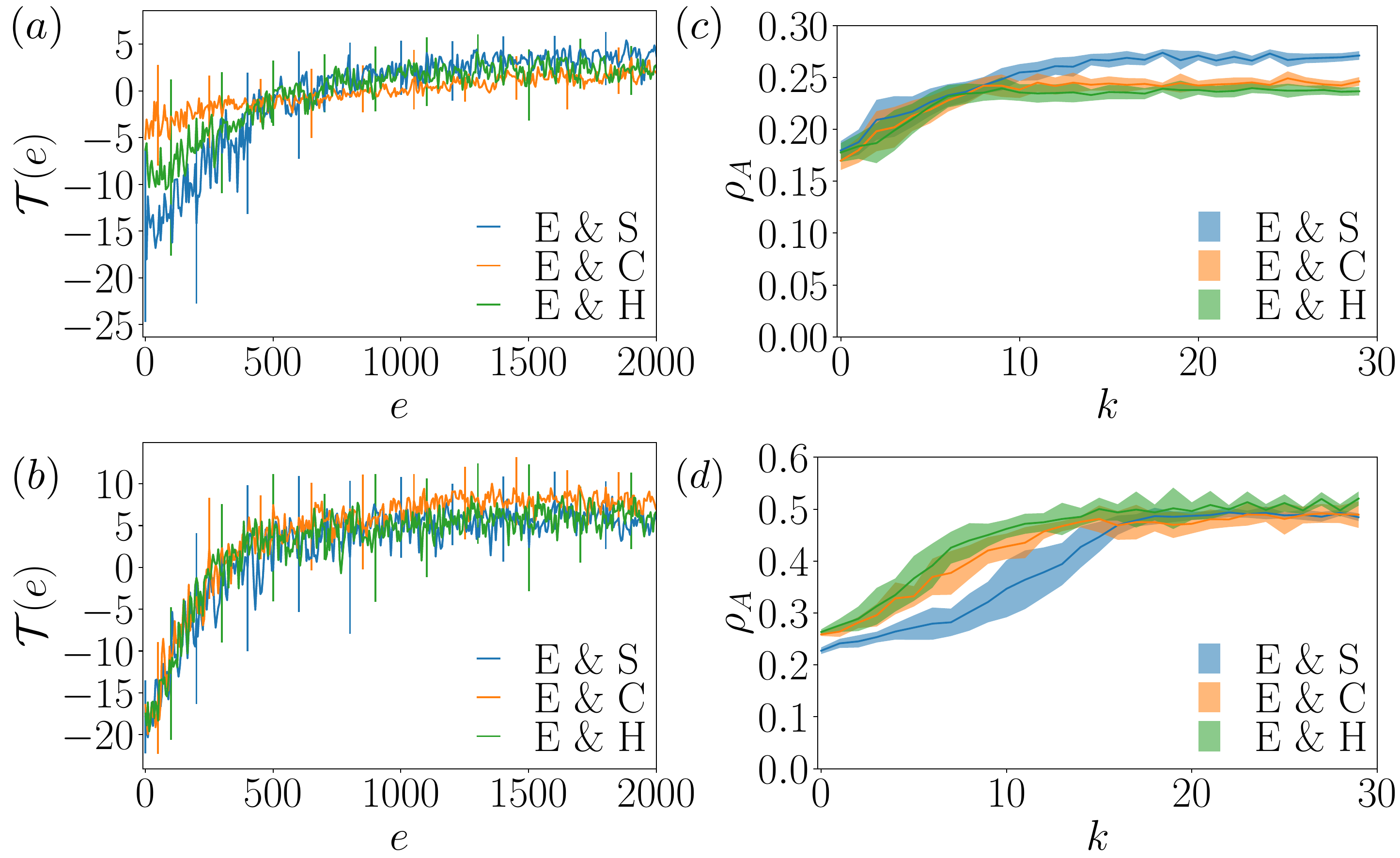}
    \caption{
    (a) Integrated reward $\mathcal{T}(e)$ for mixed strategies with $p_{\omega}=0.015$.  The data is shown as the average of $10$ runs.
    (b)~Evolution of $\rho_A$ for RL runs starting from a pre-learned value function at $(\sigma_A=1.0, \lambda_A=1.0, \phi=0)$ with $p_{\omega}=0.015$.
    (c,d) Similar data to (a,b) at $p_{\omega}=0.005$.
    }
    \label{fig:learning_curve_4d}
\end{figure}

As well as pure strategies (hunt, evade, etc), we also consider mixed strategies that combine evasion with other characteristics.  Fig.~\ref{fig:learning_curve_4d} demonstrates convergence for this four-parameter optimisation, analogous to Fig.~\ref{fig:learning_curve}.  One sees from Fig.~\ref{fig:density_compare} that for the crowded case, the optimal strategy found by RL always reverts to pure evasion ($\phi_2=0$).  In the sparse case, the mixed evasion \& spreading strategy does improve the $A$ population, but the other mixed strategies again revert to pure evasion. 
Table~\ref{tab:optimal_para} summarises the optimal state points found by RL, for the various strategies.

\subsection{Interpretation of learned strategies}
\label{sec:interpretation-rl}

We discuss the results of Fig.~\ref{fig:density_compare} and Tab.~\ref{tab:optimal_para}.  We first compare the symmetric case ($A$ behaves identically to $B,C$) with the non-directional case ($\sigma_A,\lambda_A$ are optimised but particles have no directional preferences) and we focus on the sparse situation ($p_{\omega}=0.015$).  Tab.~\ref{tab:optimal_para} shows that it is desirable for $A$ to move faster than the other species ($\lambda_A>1$) but consume less prey ($\sigma_A<1$).   We explain below (Sec.~\ref{sec:result}) that this reduced $\sigma_A$ results in more hungry particles and hence reduced reproduction rate, but this apparent reduction in fitness is counteracted by the ``survival of the weakest'' effect~\cite{frean2001rock, kerr2002local, berr2009zero, nahum2011evolution, menezes2019uneven}, which is typical for systems with cyclic dominance.  The key insight is that reduced predation by $A$ enhances the population of species $B$, and this species in turn predates on $C$, reducing their population.  Recalling that $C$ acts in turn as a predator for $A$, this effect tends to also enhance $\rho_A$.  This effect is apparent throughout Fig.~\ref{fig:density_compare} because the optimised parameters always lead to reduced $C$ populations, reducing the amount of predation on species $A$.

Comparing symmetric and non-directional strategies for the crowded case, the optimal parameters now have strongly reduced $\sigma_A$, which again facilitates survival of the weakest.  (In this situation, the optimal $\lambda_A$ is reduced with respect to the other species, which is opposite to the sparse case.)

Turning to pure directional strategies, there is a significant improvement over non-directional strategies in the sparse case, with both evasion and spreading proving effective.  (Recall that evasion corresponds to moving away from predators, while spreading corresponds to moving into empty space.)   The survival-of-the-weakest effect hints that predation plays an important role in determining $A$'s population, so it is not surprising that evasion of predators is also effective.  The role of spreading is not so clear-cut but we recall that particles can only reproduce if empty space is available, so this strategy naturally increases the net rate of reproduction.  In the crowded case, the evasion strategy provides a marginal benefit, although the population of the predator ($C$) becomes very low.  If the $C$ dies out then the ecosystem will collapse: we do observe that the $A$ population has quite large fluctuations, indicated by the error bar in Fig.~\ref{fig:density_compare}, see also Sec.~\ref{sec:result}, below.

\begin{figure}
    \centering 
    \includegraphics[width=0.49\textwidth]{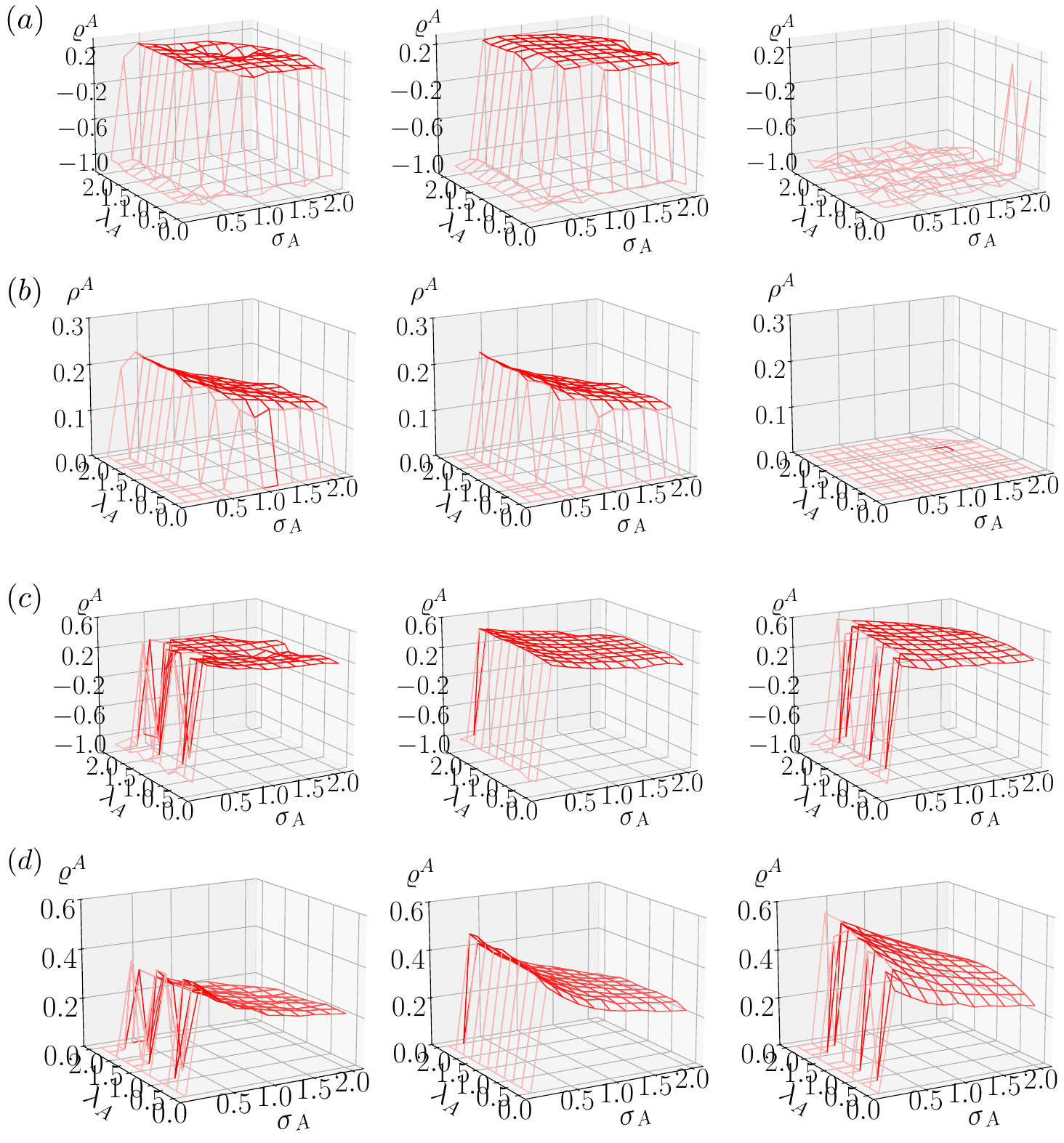}
    \caption{(a)~Learned value function $\varrho(S)$ for $p_{\omega}=0.015$ as a function of $\sigma_A$ and $\lambda_A$ at $\phi_{\rm{E}}=0, 1.25,2.5$ (left to right).  
    (b)~Population $\rho_A$ for the same parameters shown in (a).  
    (c)~Learned value function $\varrho(S)$ for $p_{\omega}=0.005$, again with $\phi_{\rm{E}}=0, 1.25,2.5$.
    (d)~Similar data for $p_{\omega}=0.005$.
    }
    \label{fig:learnt_density}
\end{figure}

Note that the strategy of hunting prey is never effective: optimal strategies always have $\phi_{\rm H}=0$. 
This can also be rationalised via survival of the weakest since hunting prey reduces the $B$ population, which allows the predator population $C$ to grow, eventually harming $A$.  The clustering strategy (movement of $A$ particles towards others of the same species) has no benefit in the sparse case but does have a positive effect in the crowded case.  This is likely due to $A$ particles shielding each other from predators (there is at most one particle per site so a high local density of $A$'s tends to reduce the density of $C$'s).

As noted above, survival-of-the-weakest achieves a large $A$ population by suppressing their predators (species $C$): however, if the $C$ population falls too low then a random fluctuation may cause them to die out, in which case the ecosystem collapses and all species become extinct.  This effect is illustrated in Fig.~\ref{fig:learnt_density} which shows both the value function $\varrho$ (estimated by RL) and the $A$ population $\rho_A$, as a function of $\sigma_A$, for various $\lambda_A,\phi_{\rm E}$.  Survival of the weakest corresponds to $\varrho,\rho_A$ decreasing with $\sigma_A$.  However, if $\sigma_A$ falls too low then species $B$ becomes very numerous and species $C$ is suppressed, leading to ecosystem collapse and $\varrho=-1$.  We note that the inclusion of hunger and natural death in the model is necessary for ecosystem collapse and extinction.  Without these effects $\sigma_A=0$ is typically the optimal parameter value~\cite{frean2001rock, kerr2002local, berr2009zero, nahum2011evolution, menezes2019uneven}.

\begin{figure*}[t!]
    \centering 
    \includegraphics[width=0.98\textwidth]{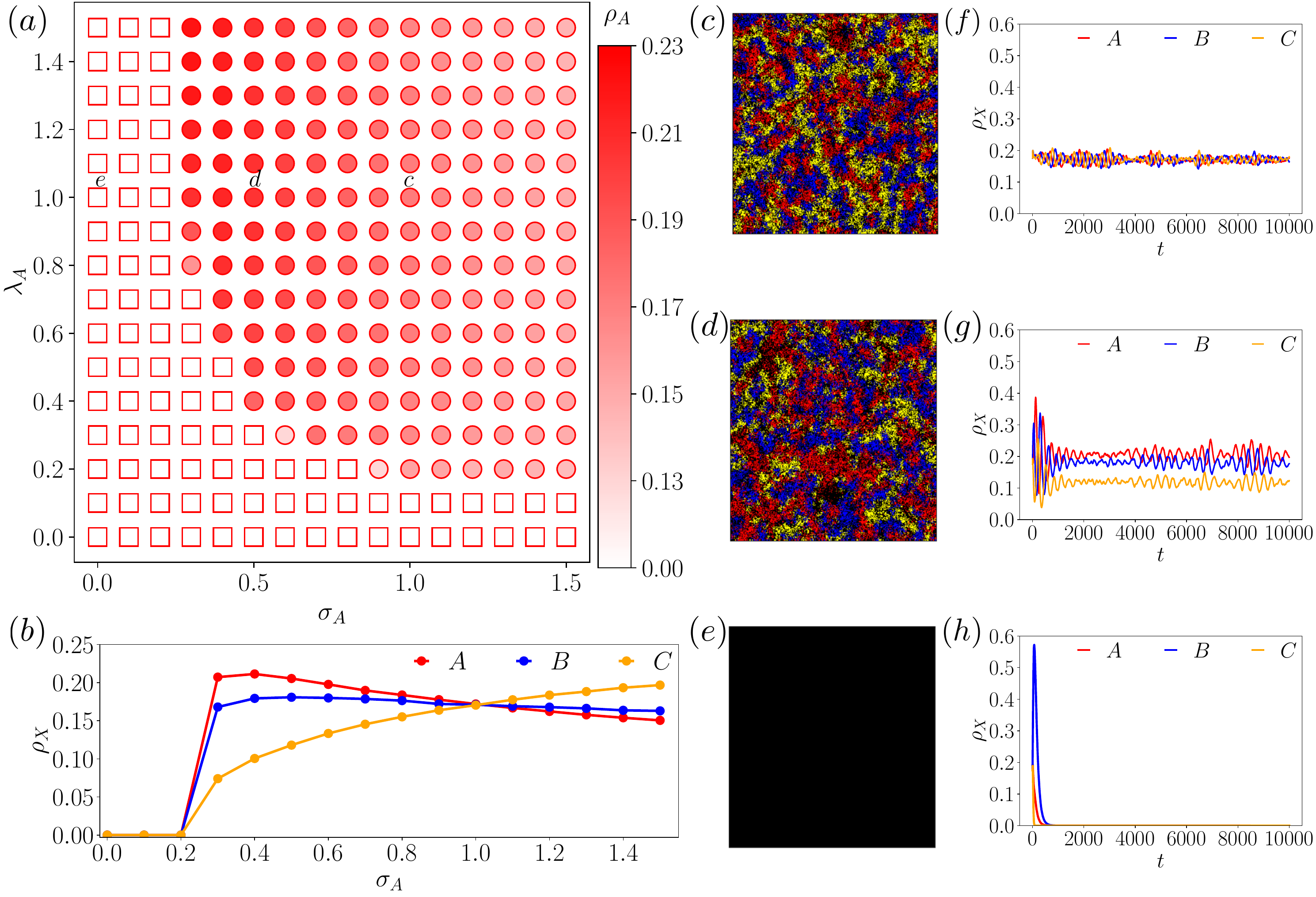}
    \caption{(a) Population density of ${A}$ particles as a function of $\sigma_{A}$ and $\lambda_{A}$ at $p_{\omega}=0.015$.  ${A}$ particles becomes extinct in low $\sigma_{A}$ and $\lambda_{A}$ region and coexist with ${\rm B, C}$ particles in the high $\sigma_{A}$ and $\lambda_{A}$ region.  Within the coexistence region, lower $\sigma_{A}$ corresponds to higher density of ${A}$ and vice versa.  (b) The density of three different types of particles as a function of $\sigma_{A}$ at $\lambda_{A}=1.0$.  (c, d, e) Three different types of behaviour of the system at $\sigma_{A}=1.0, 0.5, 0.0$ respectively, corresponding to symmetric coexistence, asymmetric coexistence where ${A}$ has increased density and extinction where all three types of particles become extinct.  All simulations are performed with $L=300$ and $T=10^5$.
    }
    \label{fig:predation_move_diagram_low_density}
\end{figure*}

\section{Physical Interpretation of Advantageous Strategies}\label{sec:result}

This Section describes in more detail the model behaviour, including the competition between species' populations and the role of hunger levels and spontaneous death processes.  We focus on $p_{\omega}=0.015$ (sparse case), the behaviour for the crowded case is provided in Appendix~\ref{sec:sft_high_density}, for comparison.

\subsection{Survival of the Weakest (non-directional movement)}\label{sec:survival_of_weakest}

To complement the results of RL, Fig.~\ref{fig:predation_move_diagram_low_density} illustrates the behaviour of the system with non-directional movement strategy, with parameter scans for $\lambda_A,\sigma_A$.  
We take $L=300$, consistent with Sec.~\ref{sec:phase_behaviour}.
Fig.~\ref{fig:predation_move_diagram_low_density}(a) shows the $A$ population density $\rho_A$, showing extinction for small $\lambda_A,\sigma_A$ (leading to $\rho_A=0$); there is a stable ecosystem for larger $\lambda_A,\sigma_A$, with $\rho_A$ decreasing with $\sigma$ due to survival of the weakest (recall Fig.~\ref{fig:learnt_density}).  Fig.~\ref{fig:predation_move_diagram_low_density}(b) shows that the $C$ population $\rho_C$ is anti-correlated with $\rho_A$.  However, as discussed in Sec.~\ref{sec:interpretation-rl}, this effect cannot continue to arbitrarily small $\sigma_A$ because species $C$ tends to die out, and the ecosystem collapses.  

To see this more clearly we identify three representative state points which have $\lambda_A=1.0$ and $\sigma_A=0.0,0.5,1.0$.  
Figs.~\ref{fig:predation_move_diagram_low_density}(c,d,e) show snapshots from these state points: the $C$-population is small in (d) which favours species $A$.
 Figs.~\ref{fig:predation_move_diagram_low_density}(f,g,h) show the time series of the species densities.  The oscillations are characteristic of cyclic dominance (and for predator-prey dynamics more generally).   For case (g) the oscillations in $C$ population are significant but the population remains always away from extinction.  For case (h) where $\sigma_A=0$ there is no predation on the $B$ species so its population grows quickly, and this results in extinction.

\begin{table}[t]
\centering
\begin{tabular}{|c|c|c|c|c|}
\hline
                                           & $\rho_X$  & $\rho_{X^0}+\rho_{X^{\prime}}$ & $\rho_{X^{\prime\prime}}$ & $\rho_{X^{\prime\prime}}/\rho_X$ \\ \hline
$A,B,C$ (sym) & $0.171$ & $0.045$                        & $0.126$                   & $0.737$                                  
\\ \hline
$A$    (asym)                        & $0.205$ & $0.047$                        & $0.158$                   & $0.771$                                  \\ \hline
$B$      (asym)                      & $0.181$ & $0.044$                        & $0.137$                   & $0.757$                                  \\ \hline
$C$                 (asym)           & $0.118$ & $0.034$                        & $0.084$                   & $0.712$                                  \\ \hline
\end{tabular}
\caption{Total density and densities separated by hunger level for null (symmetric) strategy, and for the non-symmetric strategy without directional incentive (asym) for $p_{\omega}=0.015$.
}
\label{tab:sft}
\end{table}

\begin{figure}[t]
    \centering 
    \includegraphics[width=0.48\textwidth]{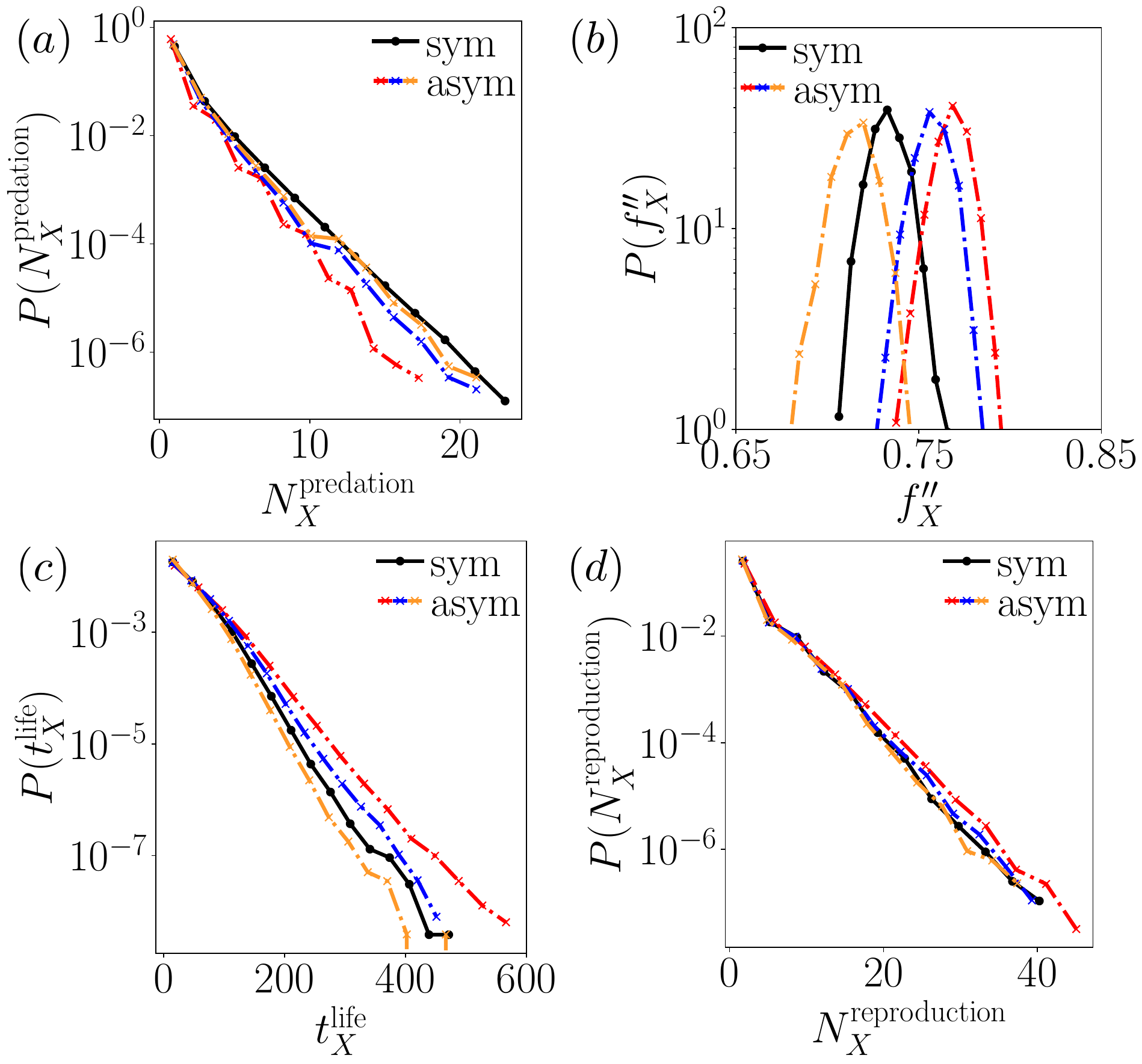}
    \caption{
    (a) The probability distribution of the predation count of individual particles. 
    (b) The probability distribution of the fraction of particles in each species having the highest hunger level.
    (c) The probability distribution of the life expectancy of individual particles. 
    (d) The probability distribution of the reproduction count of individual particles. 
    All data are collected with $L=300$ and $p_{\omega}=0.015$. The black line corresponds to the symmetric case. The coloured lines correspond to the respective types of particles with $\sigma_A=0.5$. 
    }
    \label{fig:low_density_sft}
\end{figure}

Note that $\sigma_A=1.0$ is the symmetric case where all species behave identically.  Table~\ref{tab:sft} shows a comparison of this case with the non-symmetric state point $\sigma_A=0.5$.  Specifically, the Table decomposes the steady-state populations according to their hunger level.  The non-symmetric case has the higher $A$ population, but this increase is mostly among the particles with the highest hunger level ($A''$).  These particles have a reduced reproduction rate so they contribute little to the propagation of the species: the low value of $\sigma_A$ means that they do not consume too much prey ($B$), so the $B$ population remains large, which reduces in turn the density of predators $C$.  This is how survival of the weakest operates in this model, notwithstanding the differences from previous work (that too small a value for $\sigma_A$ leads to the death of the $C$ species and hence extinction of all species).

\subsubsection{Particle Demographic Data}
\label{sec:demo-nonsym}

These data for particle hunger levels are interesting for the ecological context of this model,
because large numbers of hungry particles are optimal for the species population, even though these particles have individually lower fitness (lower reproduction rate)~\cite{bohannan2002trade, kerr2004altruism, szolnoki2014cyclic, park2020cyclic}.  Motivated by this observation, we analyse individual particles' properties in more detail.

\begin{figure}[t]
    \centering 
    \includegraphics[width=0.48\textwidth]{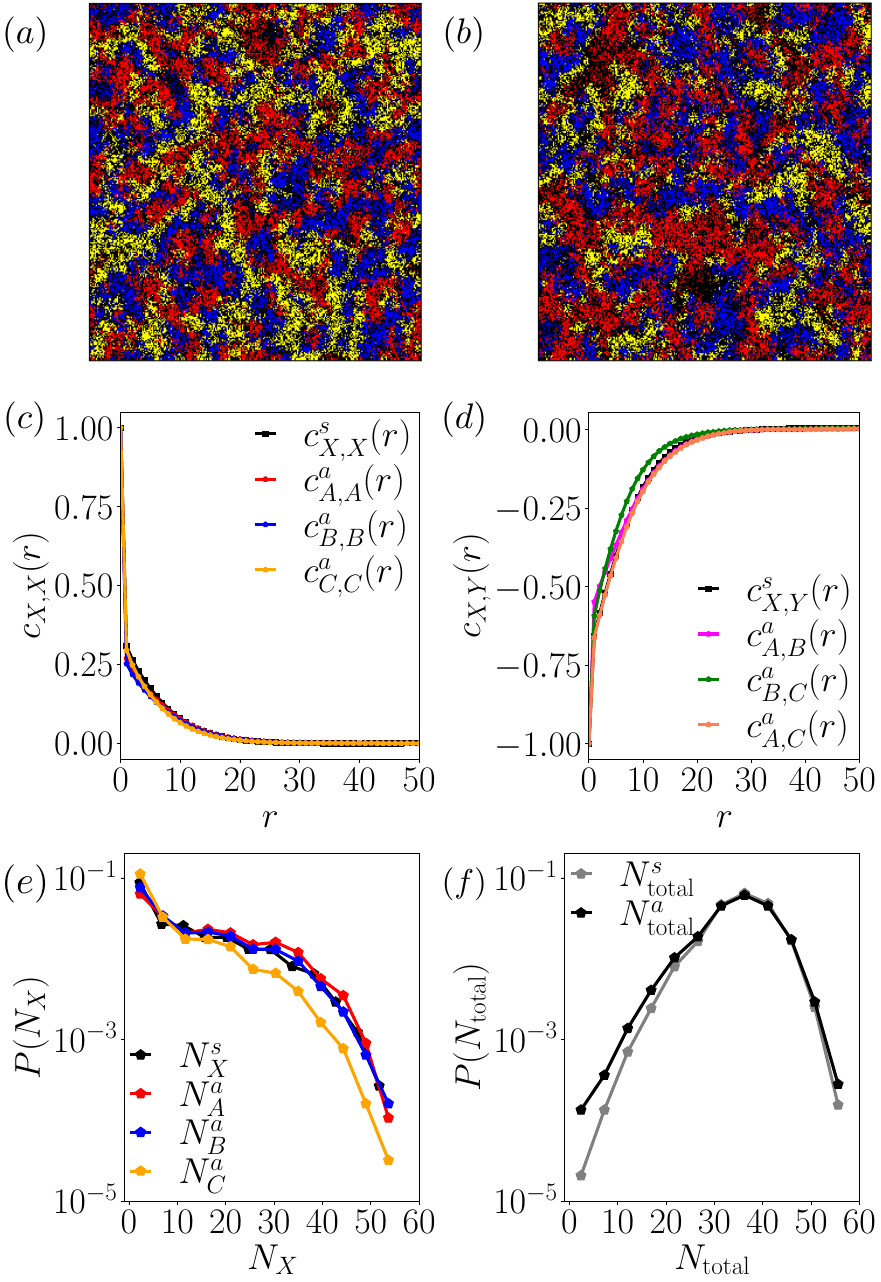}
    \caption{(a, b)~Steady-state snapshots with $\sigma_A=1$ and $\sigma_A=0.5$ respectively.  
    (c, d)~The normalised same species correlation functions $c_{X,X}(r)$ and the normalised cross species correlation functions $c_{X,Y}(r)$ (with $Y\neq X$).  Black lines show the symmetric case. Coloured lines are obtained for $\sigma_A=0.5$. 
    (e)~The distribution of the number of individual species of particles in a randomly selected circular probe region of radius $r_0=5$, the colouring is the same as panel (c).
    (f)~The distributions of the total number of particles in a randomly selected circular probe region of radius $r_0=5$.  
    All data are collected with $L=300$ and $p_{\omega}=0.015$.  
    }
    \label{fig:low_density_sft_corr}
\end{figure}

\begin{figure*}[t]
    \centering 
    \includegraphics[width=0.98\textwidth]{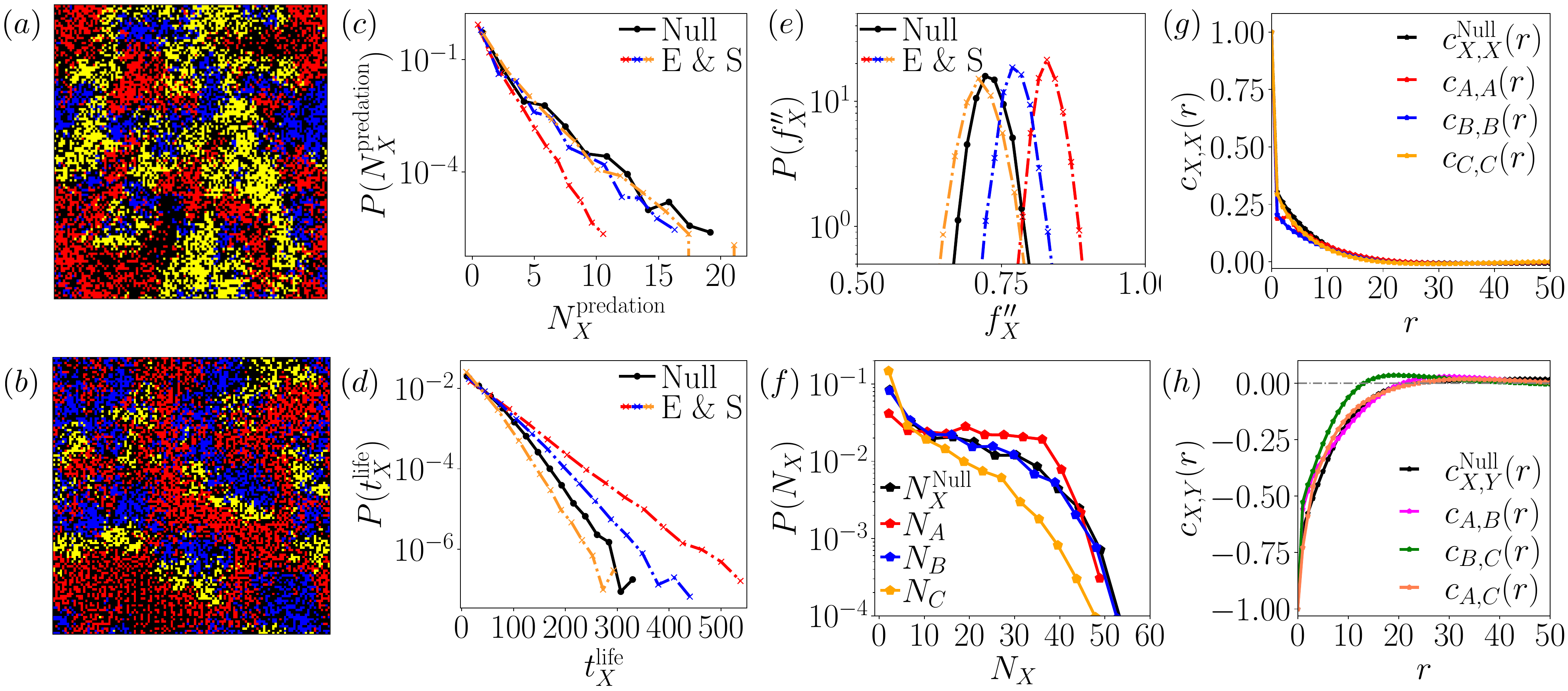}
    \caption{
    Steady-state behaviour for  $p_{\omega}=0.015$ comparing the null and evade \& spread (E\&S) strategies.
    (a, b) Snapshots from the steady state for the null and E\&S cases respectively. 
    (c, d) The normalised same-species correlation functions $c_{X,X}(r)$ and the normalised cross-species correlation functions $c_{X, Y}(r)$, coloured data are for E\&S, black for null.
    (e) The fraction of particles at their highest hunger level.
    (f) The distribution of the number of individual particles of each species in a randomly selected circular probe region of radius $r_0=5$. 
    (g) The distribution of the predation count of individual particles.
    (h) The distribution of the life expectancy of individual particles.
    All data are obtained with $L=120$, fixed parameters are given in Table~\ref{tab:fixed_para}.
    }
    \label{fig:low_density_strategy}
\end{figure*}

As discussed in Sec.~\ref{sec:model_definition}, we keep track of three particle-specific quantities: age, predation count, and reproduction count.  These quantities accumulate throughout the lifetime of individual particles.  When a particle dies, we record these quantities and collect their statistics.  For any steady state, the average reproduction count is always unity (because every particle dies exactly once, and a steady state must have the number of births matching the number of deaths).

We run simulations on $L\times L=300\times 300$ systems of $3\times 10^5$ MCS.  During the first $10^5$ MCS we allow the system to relax into its steady state.  We collect particle statistics for the following $2\times 10^5$ MCS.

Figure~\ref{fig:low_density_sft}(a,c,d) shows histograms for predation count, particle lifetime, and reproduction count.  
Fig.~\ref{fig:low_density_sft}(b) shows the histogram for the fraction $f''_X$ of particles in the highest hunger level.  (We compare the symmetric case with $(\sigma_A,\lambda_A)=(0.5,1.0)$ similar to Table~\ref{tab:sft}.)  Similar data for $p_{\omega}=0.005$ is shown in Appendix~\ref{sec:sft_high_density} for completeness.  
These results have several features. First, the non-symmetric case does indeed have reduced predation counts for $A$. Second, the fraction of $A$ particles in hunger level 2 is correspondingly increased, consistent with Table~\ref{tab:sft}; the corresponding fraction of $B$ particles is also enhanced (presumably because their prey species $C$ are suppressed).  The fraction of $C$ particles in this hunger level is reduced because their prey species $A$ is numerous.   Third, the lifetime of the $A$ particles is enhanced, which we attribute to the low population of their predators ($C$).  Similarly, $B$ is also enhanced, because their predators ($A$) have reduced predation rate $\lambda_A$.  Fourth, the distribution of reproduction counts is similar to the symmetric case, despite the different lifetimes.  (The increased lifetimes of $A,B$ are balanced by their lower net reproduction rates, which arise in turn from their higher fractions of hungry individuals.)

These results illustrate the implications of the survival-of-the-weakest effect for individuals: the privileged species are also more numerous but they also tend to be hungrier.

\subsubsection{Density Fluctuations}
\label{sec:correl-nonsym}

A striking feature of the rock-paper-scissors models is the self-organisation of species into spiral waves.  In this Section we analyse spatial correlations of the species' densities, to understand how this self-organisation differs between symmetric and non-symmetric cases. Fig.~\ref{fig:low_density_sft_corr}(a,b) shows representative snapshots of these two cases.

Recalling that $\eta_X(\bm{r})$ is the number of particles of species $X$ at position $\bm{r}$,
the normalised two-point correlation functions between particle types $X,Y$ are 
\begin{equation}
    C_{X,Y}(\boldsymbol{r},\boldsymbol{r}') = \frac{1}{\mathcal{N}_{X,Y}}(\langle\eta_X(\boldsymbol{r})\eta_Y(\boldsymbol{r}') \rangle - \langle \eta_X(\boldsymbol{r}) \rangle \langle \eta_Y(\boldsymbol{r}') \rangle),
\end{equation}
where the normalisation factor is $\mathcal{N}_{X,X}=\langle \rho_X \rangle ( 1 - \langle \rho_X \rangle)$ while $\mathcal{N}_{X,Y}=\langle \rho_X \rangle\langle \rho_Y \rangle$ for $X\neq Y$.
This normalisation means that $C_{X,Y}$ reveals the spatial structure of the correlations, independent of the species' average densities.
When presenting numerical results, we use superscripts on $c_{X,Y}$ to indicate the strategy used, for example $c^s$ for symmetric (null) strategy and $c^a$ for the asymmetric (but non-directional) strategy.

\begin{figure*}[t]
    \centering 
    \includegraphics[width=0.98\textwidth]{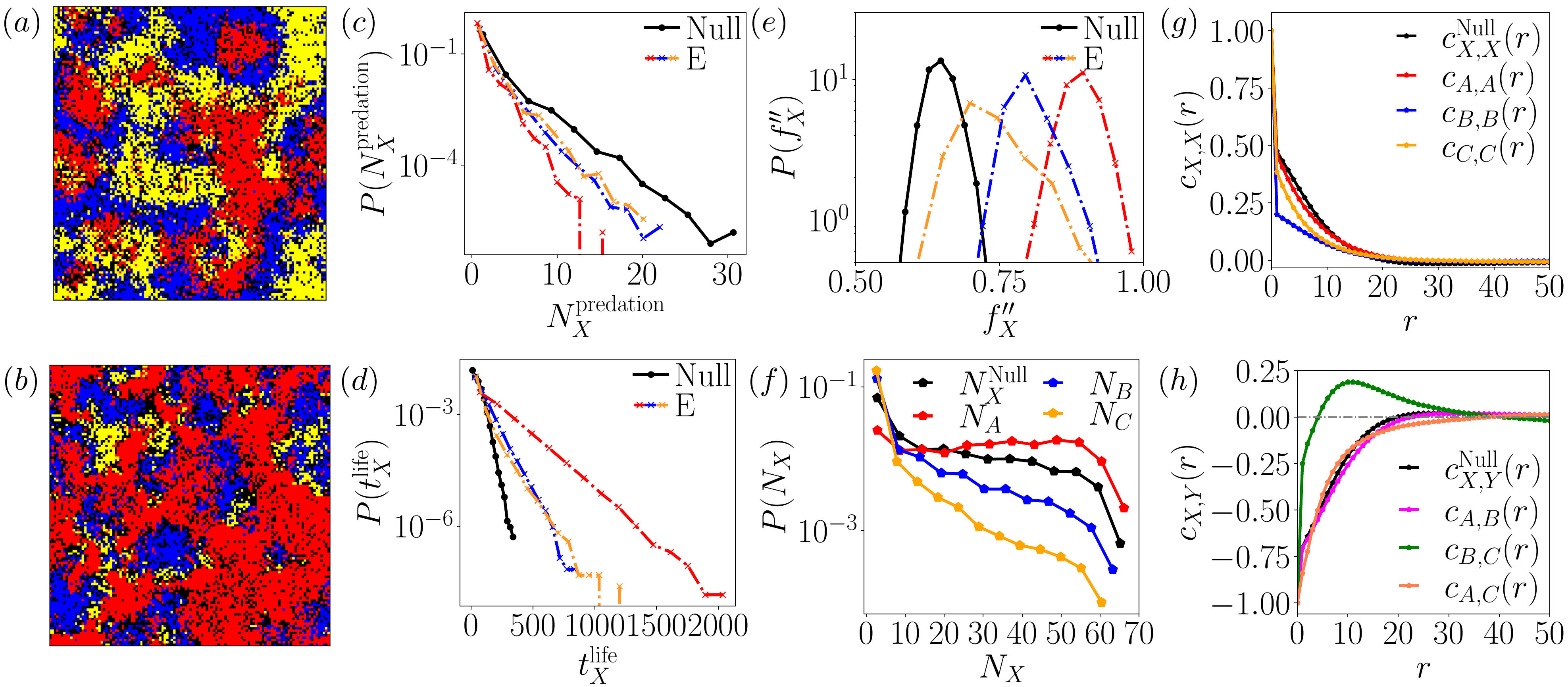}
    \caption{
    Steady-state behaviour for  $p_{\omega}=0.005$ comparing the null and evade (E) strategies.
    (a, b) The snapshots of null strategy and evade strategy respectively.
    (c, d) The normalised same-species correlation functions $c_{X,X}(r)$ and the normalised cross-species correlation functions $c_{X, Y}(r)$.
    (e) The fraction of particles at their highest hunger level.
    (f) The distribution of the number of individual particles of each species in a randomly selected circular probe region of radius $r_0=5$. The black curve is obtained with the null strategy. 
    (g) The distribution of the predation count of individual particles.
    (h) The distribution of the life expectancy of individual particles.
    All data are obtained at $L=120$, fixed parameters are given in Table~\ref{tab:fixed_para}.
    }
    \label{fig:high_density_strategy}
\end{figure*}

These correlations were estimated using simulations of $10^5$ MCS to ensure the system reaches a steady state and collecting data over the next $2\times 10^5$ MCS.  Results are shown in Fig.~\ref{fig:low_density_sft_corr}(c,d).  The `self' correlations $c_{X,X}$ all behave similarly, showing clustering of all species over similar length scales, of the order of $10$ lattice spacings.   

The correlations between species are negative, indicating an effective repulsion: this is expected from the combination of the clustering and the exclusion constraint but it is also affected by predation.  For example, the fact that the non-symmetric case has less predation of $B$ by $A$ means that $c_{A,B}$ is less negative at short distances, compared with symmetric case.  The non-symmetric case also has $c_{B,C}$ less negative at intermediate distances (for example, $r\approx 10$), indicating a change in the arrangement of the patches of different species.  See Sec.~\ref{sec:es_results} for further discussion of this effect.

A complementary measure of clustering is obtained by choosing a random circular area in the system and counting the number of each species within that ``probe'' area~\cite{crooks1997gaussian, del2018energy, omar2021phase, yu2024competition}  We take circles of radius $r_0=5$, comparable with the cluster size inferred from the two point correlations.  Results for individual species are shown in Fig.~\ref{fig:low_density_sft_corr}(e).  From the tail of the histogram, we observe an increased probability of having no $C$ particles at all, and a decrease in the number of large clusters of the $C$ species (we attribute this to the reduced $C$ density.)  Fig.~\ref{fig:low_density_sft_corr}(f) shows the distribution of the total number of particles in the probe area.  In the non-symmetric case, there is a significant increase in the probability to have very few particles in the probe area, {which we again attribute to the reduced $C$ density, which promotes larger fluctuations. (For example, $C$ is more likely to die out locally, causing their prey $A$ to proliferate, until such time as predators arrive some elsewhere and control them.)}

Overall, we find mild differences between symmetric and non-symmetric cases, as one may expect because particles' interactions still have random directions, even if one species behaves differently from the others.  In the following we discuss some effects of directional strategies.

\subsection{Evade and Spread strategy (sparse case)}\label{sec:es_results}

We consider the sparse case ($p_{\omega}=0.015$) in which the Evade and Spread strategy leads to the largest population of $A$ particles (recall Fig.~\ref{fig:density_compare}).  The results of this section have $L=120$, consistent with Sec.~\ref{sec:learning}.  The relevant demographic analysis and spatial structures are characterised in Fig.~\ref{fig:low_density_strategy}.  The parameters are those of Tab.~\ref{tab:optimal_para}, note in particular that $(\sigma_A,\lambda_A)$ have the same values as the non-symmetric strategy of Secs.~\ref{sec:demo-nonsym} and \ref{sec:correl-nonsym}, but the E\&S strategy means that particles also have significant directional preferences.

We compare in Fig.~\ref{fig:low_density_strategy}(a,b) the behaviour of the Evade and Spread (E\&S) strategy with the symmetric (null) case.  As well as the more numerous $A$ particles in the E\&S case, one also sees in Fig.~\ref{fig:low_density_strategy}(b) that $A$ particles tend to be more spread out inside their domains, due to spreading.

Fig.~\ref{fig:low_density_strategy}(c) shows that the predation count of $A$ particles is significantly reduced by E\&S.  This is expected because of the reduced $\sigma_A$, but the effect is much stronger than Fig.~\ref{fig:low_density_sft}, presumably because the spreading strategy causes $A$ particles to move away from their prey species $B$.
Fig.~\ref{fig:low_density_strategy}(d) shows that the lifetime of $A$ particles is enhanced.  Again, this is a stronger version of the effect shown in Fig.~\ref{fig:low_density_sft}, which we attribute to $A$ evading their predators, and hence living longer.  The life expectancy of $B$ particles also sees a moderate increase, which we also attribute to the biased movement of $A$ away from $B$, due to spreading (recall that $A$ is the predator of $B$).
Figs.~\ref{fig:low_density_strategy}(e,f) shows the fraction of particles in the highest hunger level and the number of particles in circular probe areas.  The results are similar to Figs.~\ref{fig:low_density_sft}(b) and \ref{fig:low_density_sft_corr}(e) for the non-directional case, but again the effect is stronger.

Fig.~\ref{fig:low_density_strategy}(g,h) show correlation functions [comparable with \ref{fig:low_density_sft_corr}(c,d)].  An interesting feature is that $c_{B,C}(r)$ is positive for intermediate distances $r\gtrsim15$ and decays to zero from above as $r\to\infty$.  {To understand this, note that the movement process (Fig.~\ref{fig:model_definition}) allows particles to swap places. Suppose that $A$ and $B$ are neighbours which swap positions: then the bias for $A$ to move away from $C$ means that $B$ is biased towards $C$.  
This leads to $c_{B,C}>0$ on these intermediate length scales.
 Comparing Fig.~\ref{fig:low_density_strategy}(g,h) with Fig.~\ref{fig:low_density_sft_corr}(c,d), one also sees that the $A,A$ correlation is reduced by E\&S for small distances ($r\lesssim3$) due to the spreading.}

Summarising, the E\&S strategy is advantageous for species $A$ because evading predators increases life expectancy.  Spreading is advantageous because movement into empty space increases the rate of reproduction (which requires an empty adjacent site, recall Sec.~\ref{sec:interpretation-rl}).  These strategies increase $A$'s life expectancy without impacting the population of their prey $B$, so that this species ($B$) continues to predate on $C$, controlling their population and reducing their ability to predate on $A$.  Note that these strategies benefit the entire population of $A$ and they also benefit individual particles, via increased lifetime.  However, each $A$ particle predates less, leading to higher hunger levels than one finds with the non-directional (or symmetric) strategies.

\subsection{Evade strategy (crowded case)}\label{sec:e_results}

For the crowded case ($p_{\omega}=0.005$), the evade strategy is optimal [see Fig.~\ref{fig:density_compare}].
This case is analysed in Fig.~\ref{fig:high_density_strategy} (again for $L=120$). This is comparable with Fig.~\ref{fig:low_density_strategy}, as we now discuss.
Figure.~\ref{fig:low_density_strategy}(a,b) compares steady-state snapshots of the null (symmetric) strategy and the evade (E) strategy.  One clearly sees that evasion leads to a large $A$ population, with very few of their predators ($C$).  
%
%
The predation count and the life expectancy for the evade strategy are shown in Fig.~\ref{fig:high_density_strategy}(c, d).  The predation count of $A$ particles is slightly reduced while their life expectancy is significantly increased (due to the small number of their predators).  
Note that particles' lifetimes are limited by the spontaneous death process, so $P(t_X^{\rm lifetime}) $ should decay at least as fast as $P(t_X^{\rm lifetime}) \sim {\rm e}^{-p_\omega t_X^{\rm lifetime}}$ at large times.  The data are close to this limit, indicating that predation by $C$ plays a relatively small role, consistent with the low $C$ population [recall Fig.~\ref{fig:density_compare}(b)].  On the other hand, the null strategy has a faster-decaying tail, indicating that predation is important.

The lifetime of $B$ is also increased with respect to the symmetric (null) case (more precisely, the large lifetime tail is enhanced).  This is presumably caused by the reduced predation rate by $A$ (note $\sigma_A=0.4$).  The lifetime distribution of $C$ has a similar tail, which we attribute to the relatively low population of its predator species $B$ [recall again Fig.~\ref{fig:density_compare}(b)].  Indeed, comparing strategies E and S in Fig.~\ref{fig:density_compare}(b), we observe that the $A$ population is affected similarly by the directional incentive, but the $B,C$ populations are lower for the $E$ strategy.  (This effect is particularly pronounced for $B$.) %

Fig.~\ref{fig:high_density_strategy}(e) shows the fractions of particles in the highest hunger level: we find that all species are hungrier when the A particles evade their predator.  The reasons seem to be different for each species: the small $\sigma_A$ tends to increase the hunger level of $A$; the small numbers of $C$ mean that $B$ struggle to find prey ($C$); the evasion of $C$ by $A$ means that $C$ struggle to find their prey ($A$).
The larger fluctuations in $f''_C$ are presumably due to their lower overall population.  
The high hunger levels and long lifetimes together reflect that particles tend to segregate into groups of their own species, which reduces both the opportunities and the risks associated with predation.
Fig.~\ref{fig:high_density_strategy}(f) shows distributions of the particle number in circular probe areas.  Interestingly, the $A$ distribution shows a local maximum at $N_A\approx 50$, which is partly attributable to the large $A$ population, but also indicates strong clustering among these particles.

Spatial correlations are shown in Fig.~\ref{fig:high_density_strategy}(g,h).  The $B,C$ correlation is again positive for intermediate-to-large distances recall Fig.~\ref{fig:low_density_strategy}(h) for the sparse case, the reason is presumably the same but the effect is even stronger in this case because it is more likely that $A$ and $B$ particles are adjacent and swap places during movement.

The emerging picture is the usual one for the survival of the weakest: species $A$ directly evades its predator species ($C$) but it also acts to control its population by maintaining a large $B$ population (because they are the predators for $C$).  This leads to $A$ particles being hungrier but living longer.  

\section{Conclusion}\label{sec:conclusion}

\begin{figure*}[t!]
    \centering 
    \includegraphics[width=0.98\textwidth]{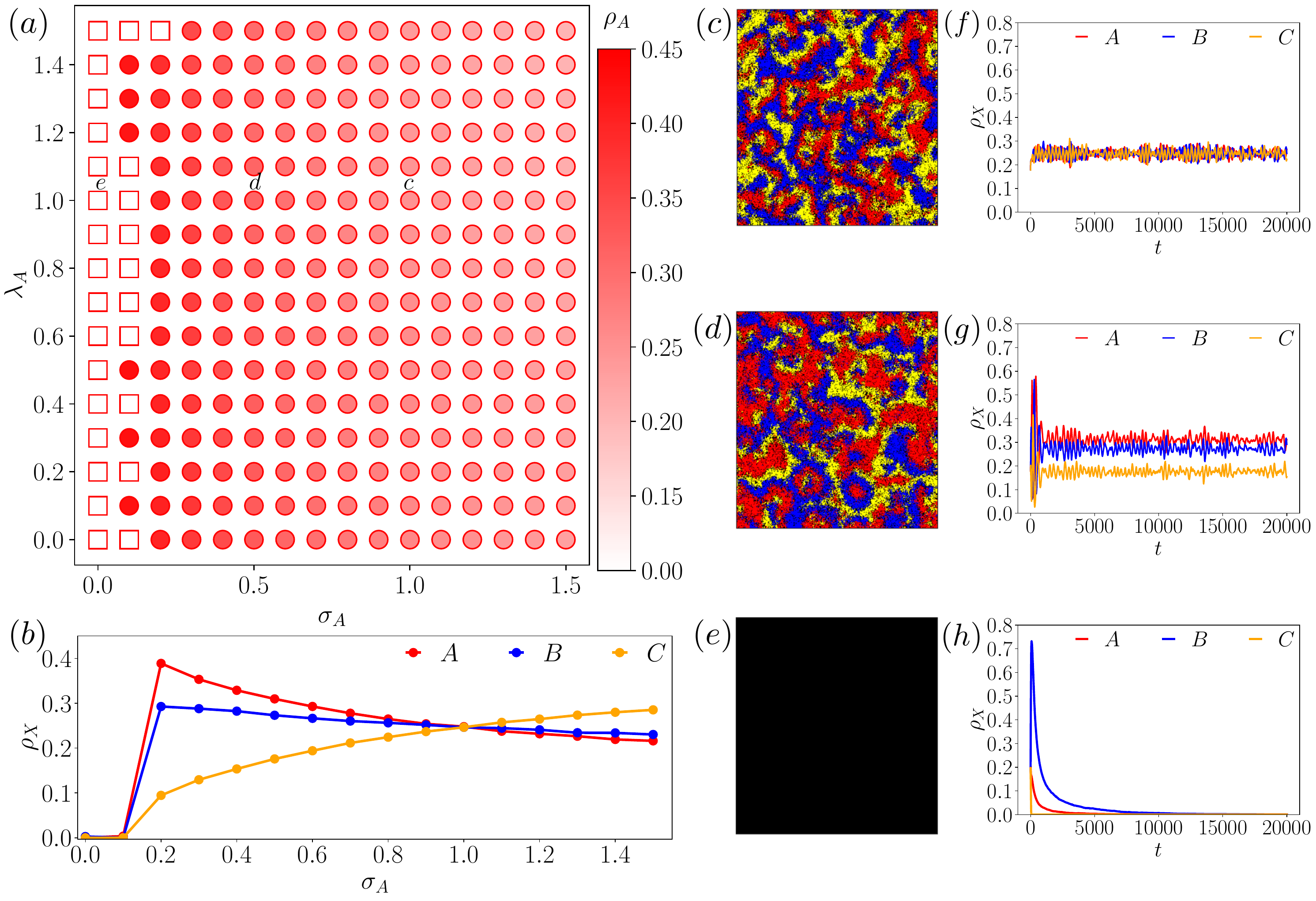}
    \caption{
    Results similar to Fig.~\ref{fig:predation_move_diagram_low_density}, but now for $p_{\omega}=0.005$.  (a) Population density diagram of ${A}$ particles as a function of $\sigma_{A}$ and $\lambda_{A}$.  (b) The density of three different types of particles as a function of $\sigma_{A}$ at $\lambda_{A}=1.0$.  (c, d, e) Three different types of behaviour of the system at $\sigma_{A}=1.0, 0.5, 0.0$ respectively.  All simulations are performed with $L=300$ and $T=10^5$.
    }
    \label{fig:predation_move_diagram_high_density}
\end{figure*}

This work generalised the rock-paper-scissors model of~\cite{reichenbach2007mobility}, with the result that individual species can only survive as part of a biodiverse state in which all three species are present.   This was achieved by incorporating hunger levels and spontaneous death processes.  We then made the further generalization that a privileged (``smart'') species ($A$) can adjust its behaviour to optimise its population.  Effective strategies for this optimisation rely on the survival of the weakest effect~~\cite{frean2001rock, berr2009zero}, in which the smart species maintains a large population of its prey, which in turn reduces the population of predators for $A$.
An interesting analogy for this effect is based on a human-tree-desert ecosystem:  by planting trees, or at least preserving trees, humans can constrain the encroachment of the desert and enhance their survivability in the ecosystem.  Even though planting trees can at some level reduce the well-being of humans such as reducing the area available as farmland, the planting of trees benefits humans overall.

The smart species additionally adopts strategies with directional incentives, for example to hunt prey or evade predators.  Using reinforcement learning to identify effective strategies, we found that evasion of $A$'s predators tends to enhance its population, as can spreading into empty space, if the system is not too crowded.  On the other hand, the survival-of-the-weakest effect explains why hunting prey is not effective in this regard.

These results raise new questions regarding the adaptability of individual species in a cyclic dominance system.  For example, the reward being optimised involves a balance between the risk of extinction and the size of the species' population.  This balance depends on the time $T$ and the penalty for extinction that appears in \eqref{eq:rt}.  It would be interesting to investigate this balance in more detail, for example by including a much larger penalty for extinction so that the species' main aim is to avoid this (catastrophic) rare event instead of optimising its population for the typical case.

Other interesting questions arise if more than one species becomes ``smart'' (able to optimise its own parameters).  One can also imagine more complex interactions among large numbers of species, in which case even richer behaviour might emerge~\cite{brown2019dynamically, szolnoki2021cooperation, park2023competition}.  Finally, we note that we have adopted the perspective of centralised learning, where the parameters for the whole species are adjusted based on its average behaviour.  An alternative perspective would treat each particle as an agent with its own learning capacity, which introduces yet more complexity to the optimization and learning processes~\cite{wang2020reinforcement, yamada2020evolution, park2021co}.  We look forward to future works in these directions.

\begin{acknowledgments}
We thank Ellery Gopaoco, Daan Frenkel, Nir Gov, Paddy Royall, Aleks Reinhardt, and Samuel W. Coles for helpful discussions.
\end{acknowledgments}

\appendix

\section{Survival of the weakest at low natural death}\label{sec:sft_high_density}
\begin{figure}[t!]
    \centering 
    \includegraphics[width=0.48\textwidth]{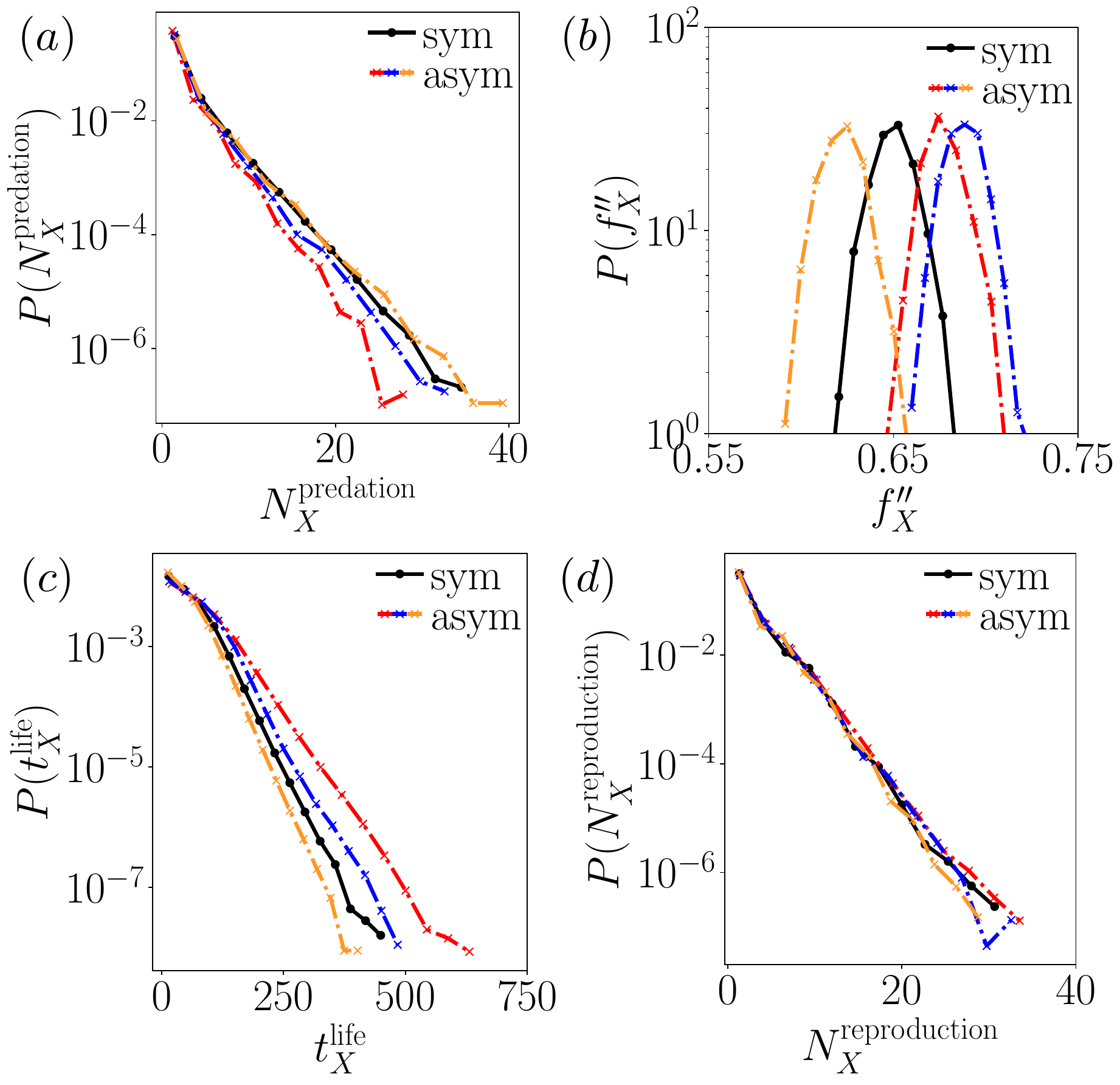}
    \caption{
    Results similar to Fig.~\ref{fig:low_density_sft}, but now for $p_{\omega}=0.005$.
    (a) The probability distribution of the predation count of individual particles. 
    (b) The probability distribution of the fraction of starving particles in each species.
    (c) The probability distribution of the life expectancy of individual particles. 
    (d) The probability distribution of the reproduction count of individual particles with at least one descendant.   System size $L=300$, other fixed parameters are given in Tab.~\ref{tab:fixed_para}.
    }
    \label{fig:high_density_sft}
\end{figure}

\begin{figure}[t!]
    \centering 
    \includegraphics[width=0.48\textwidth]{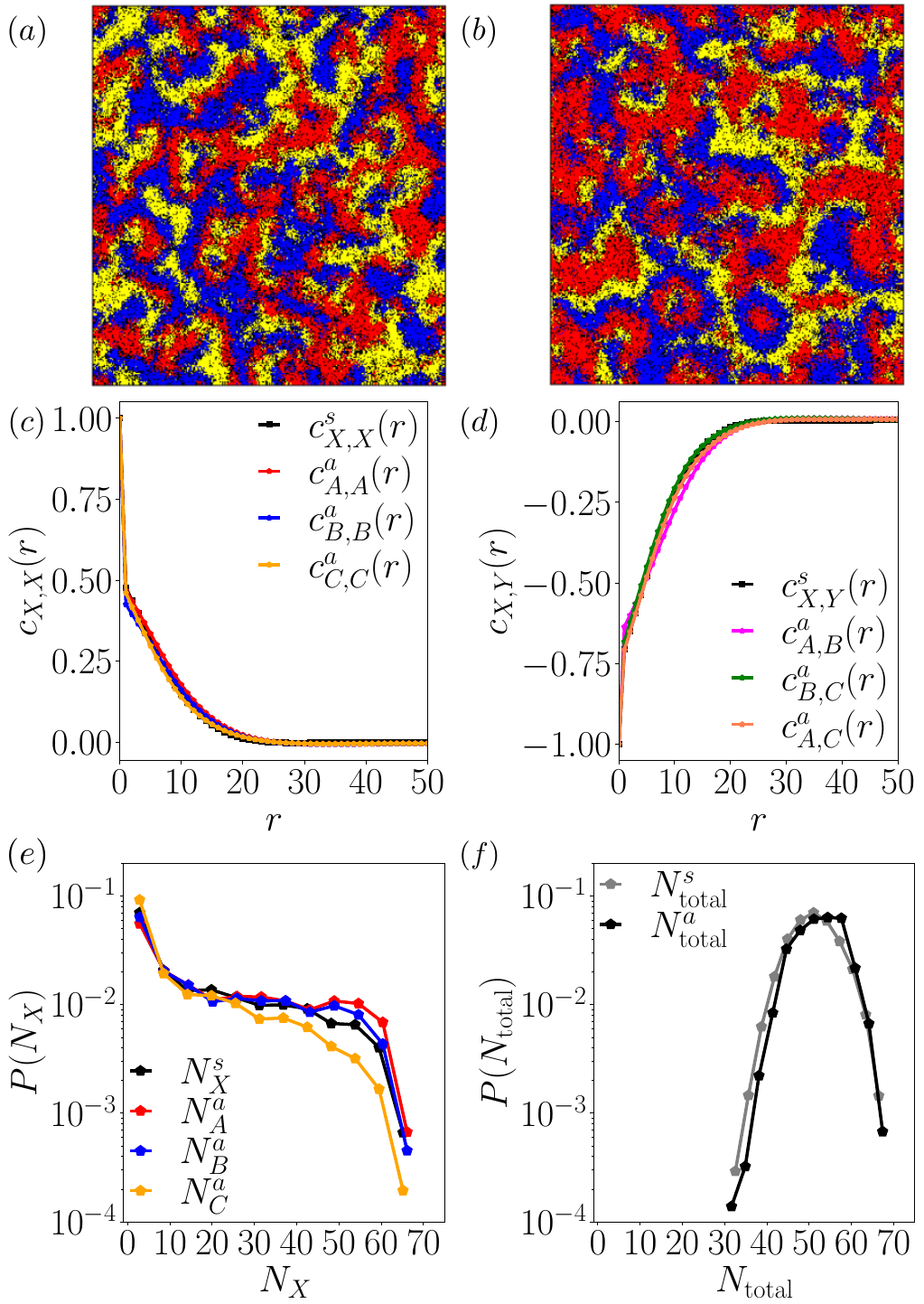}
    \caption{
    Results similar to Fig.~\ref{fig:low_density_sft_corr}, but now for $p_{\omega}=0.005$.
    (a, b)~Steady-state snapshots with $\sigma_A=1$ and $\sigma_A=0.5$ respectively.  
    (c, d)~The normalised same species correlation functions $c_{X,X}(r)$ and the normalised cross species correlation functions $c_{X,Y}(r)$ (with $Y\neq X$).
    (e)~The distribution of the number of individual species of particles in a randomly selected circular probe region of radius $r_0=5$.
    (f)~The distributions of the total number of particles in a randomly selected circular probe region of radius $r_0=5$.  
     System size $L=300$, other fixed parameters are given in Tab.~\ref{tab:fixed_para}
    }
    \label{fig:high_density_sft_corr}
\end{figure}

To complement the discussion in Sec.~\ref{sec:survival_of_weakest}, we show the survival of the weakest phenomenon in the crowded case with $p_{\omega}=0.005$.  The population density diagram of $A$ with $p_{\omega}=0.005$ is shown in Fig.~\ref{fig:high_density_sft}.  The features are consistent with the behaviour with $p_{\omega}=0.015$ as discussed in the main text.  Particle demographic data and spatial correlations at $p_{\omega} = 0.005$ as shown in Fig.~\ref{fig:high_density_sft} and Fig.~\ref{fig:high_density_sft_corr}.  Again, the general behaviour is similar to the case at $p_{\omega} = 0.015$.  We note some differences between the high and low $p_{\omega}$ cases.  At $p_{\omega} = 0.005$, the predation count is higher than the $p_{\omega} = 0.015$ case as higher particle density allows more predation as shown in Fig.~\ref{fig:high_density_sft}(a).  Recall from Fig.~\ref{fig:low_density_sft}(b), at $p_{\omega} = 0.015$, species $A$ has the highest percentage of highest hunger level particles.  However, at $p_{\omega} = 0.005$, species ${\rm B}$ has the highest percentage of highest hunger level particles among the three species.  The negative correlation $c^a_{B,C}(r)$ is weaker for $p_{\omega}=0.005$ compared to $p_{\omega}=0.015$ case.

\section{Mean-Field argument for the survival of the weakest}

The survival of the weakest phenomenon can be understood via a simple mean-field argument, following~\cite{reichenbach2008self, bhattacharyya2020mortality}. To simplify the analysis, we include natural death but do not consider the hunger mechanism. 

\begin{figure*}[t!]
    \centering 
    \includegraphics[width=0.95\textwidth]{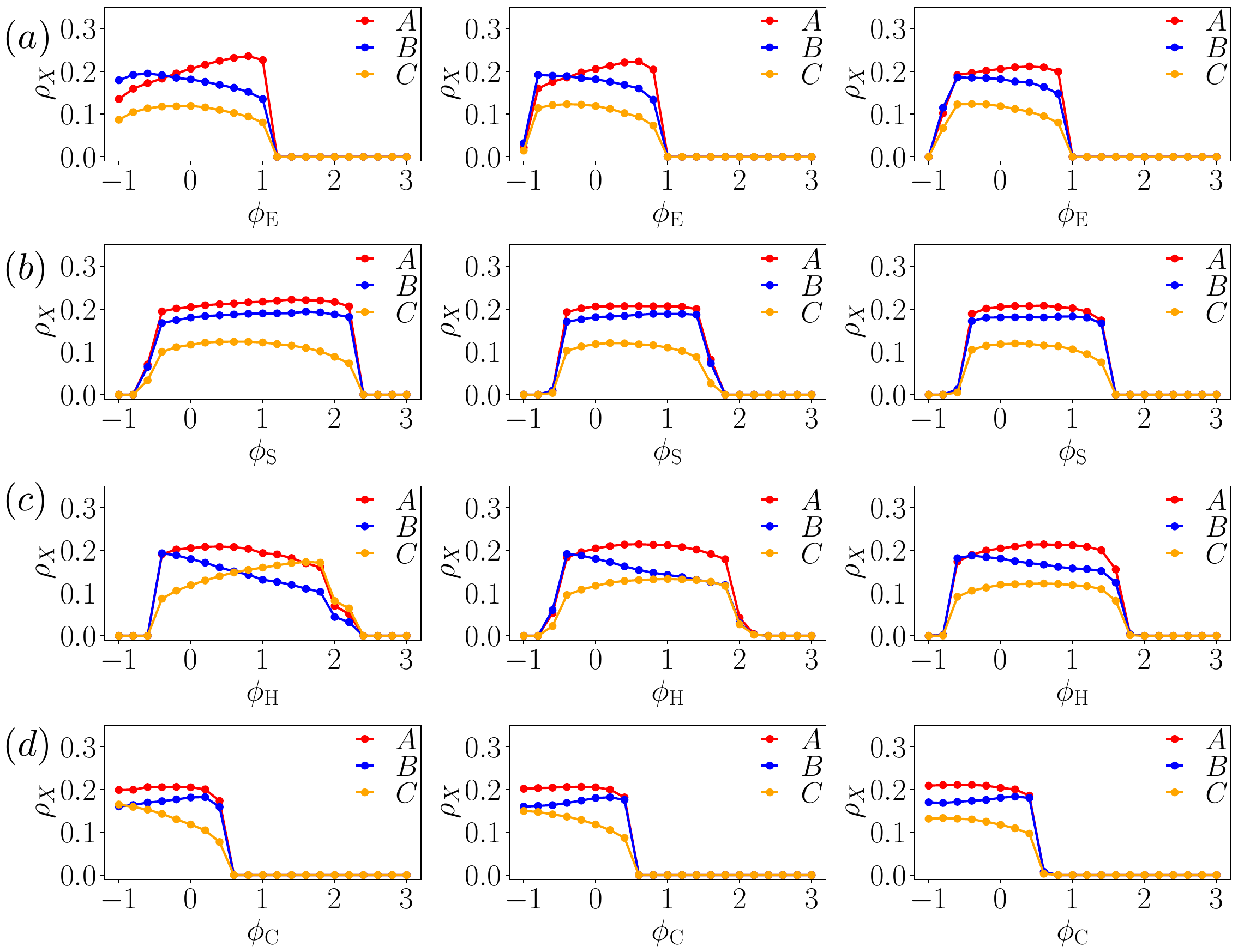}
    \caption{
    Population densities as a function of directional biases $\phi$ at $\sigma_A = 0.5$ and $\lambda_A=1.0$ for $p_{\omega}=0.015$ and four different strategies: (a) evasion, E; (b) spreading, S; (c) hunting, H; (d) clustering, C. 
    The three columns show the data obtained with $\mathcal{R}= 5, 10, 20$ respectively, other fixed parameters are given in Tab.~\ref{tab:fixed_para}.  
    The system size was $L=300$ and results were evaluated at time $T=10^5$.  
    }
    \label{fig:range_scan}
\end{figure*}

The mean density for species $X$ at position $\bm{x}$ is obtained by averaging the occupation $\eta_X$ as 
$\hat\rho(\bm{x},t) = \langle \eta_X(\bm{x},t) \rangle$ where the brackets indicate an average over many trajectories (not necessarily in the steady state of the system).  Starting from the master equation for the system's stochastic dynamics, we make two approximations~\cite{reichenbach2008self,dobramysl2018stochastic}: that $\hat\rho$ depends smoothly on $\bm{x}$, and that two-point correlations may be factorised for $\bm{x}\neq\bm{x}'$ as 
\begin{equation}
    \langle \eta_i(\boldsymbol{x}, t) \eta_Y(\boldsymbol{x}^{\prime}, t) \rangle \approx \langle \eta_X(\boldsymbol{x}, t)\rangle \langle \eta_Y(\boldsymbol{x}^{\prime}, t) \rangle.
\label{eq:MFT_approximation}
\end{equation}
which corresponds to a well-mixed (or mean-field) assumptions.

The resulting equations of motion are~\cite{reichenbach2008self,dobramysl2018stochastic}:
\begin{equation}
\begin{aligned}
    \frac{\partial \hat\rho_A(\boldsymbol{x}, t)}{\partial t}=D\nabla^2 \hat\rho_A(\boldsymbol{x}, t) + \mu_A\hat\rho_A(\boldsymbol{x}, t)\hat\rho_{\varnothing}(\boldsymbol{x}, t) \\ - 
    \sigma_C\hat\rho_A(\boldsymbol{x}, t)\hat\rho_C(\boldsymbol{x}, t) - \omega_A\hat\rho_A(\boldsymbol{x}, t),
    \\
    \frac{\partial\hat\rho_B(\boldsymbol{x}, t)}{\partial t}=D\nabla^2 \hat\rho_B(\boldsymbol{x}, t) + \mu_B\hat\rho_B(\boldsymbol{x}, t)\hat\rho_{\varnothing}(\boldsymbol{x}, t) \\ -
    \sigma_A\hat\rho_B(\boldsymbol{x}, t)\hat\rho_A(\boldsymbol{x}, t) - \omega_B\hat\rho_B(\boldsymbol{x}, t),
    \\
    \frac{\partial\hat\rho_C(\boldsymbol{x}, t)}{\partial t}=D\nabla^2 \hat\rho_C(\boldsymbol{x}, t) + \mu_C\hat\rho_C(\boldsymbol{x}, t)\hat\rho_{\varnothing}(\boldsymbol{x}, t)  \\ -
    \sigma_B\hat\rho_C(\boldsymbol{x}, t)\hat\rho_B(\boldsymbol{x}, t) - \omega_C\hat\rho_C(\boldsymbol{x}, t),
\end{aligned}
\label{eq:MFT}
\end{equation}
where we introduced $\hat\rho_{\varnothing} = 1 - \hat\rho_A -\hat\rho_B -\hat\rho_C$, for compactness of notation.
On the right-hand sides of \eqref{eq:MFT}, we identify terms corresponding to diffusion (proportional to diffusion constant $D$); reproduction (proportional to the species own reproduction rate $\mu_X$); predation (proportional to their predator's selection rate ``$\sigma_{X-1}$''); and spontaneous death (proportional to $\omega_X$).

Previous studies suggest the spatial fluctuations do not affect the qualitative behaviour of the system~\cite{reichenbach2007mobility, reichenbach2007noise, reichenbach2008self}, so we drop the spatial dependence for simplicity and introduce notation $\rho_X(t) = \hat\rho_X(\bm{x},t)$.
We obtain a system of ODEs: 
\begin{equation}
\begin{aligned}
    \frac{{\rm d} \rho_A(t)}{{\rm d} t}=\mu_A\rho_A(t)\rho_{\varnothing}(t) -\sigma_C\rho_A(t)\rho_C(t) - \rho_A(t)\omega_A,
    \\
    \frac{{\rm d} \rho_B(t)}{{\rm d} t}=\mu_B\rho_B(t)\rho_{\varnothing}(t) -\sigma_A\rho_B(t)\rho_A(t) - \rho_B(t)\omega_B,
    \\
    \frac{{\rm d} \rho_C(t)}{{\rm d} t}=\mu_C\rho_C(t)\rho_{\varnothing}(t) -\sigma_B\rho_C(t)\rho_B(t) - \rho_C(t)\omega_C,
\end{aligned}
\label{eq:MFT_ode}
\end{equation}

In general, these equations support 5 fixed points (which are solutions to $\frac{{\rm d} \rho_X}{{\rm d} t} = 0$).  One of these represents extinction ($\rho_A=\rho_B=\rho_C=0$) and there are three more that correspond to fixation.  That is, fixation of species $A$ corresponds to $\rho_A=1-(\omega_A/\mu_A)$ with $\rho_B=\rho_C=0$; the other cases are obtained by permuting the species. If the death rate $\omega_X>\mu_X$ then the associated fixed point has negative density which means that fixation of species $X$ is not possible.  

The remaining fixed point corresponds to coexistence of all three species, which is the state of primary interest in this Section.  We denote the fixed point by $(\rho_A^*,\rho_B^*,\rho_C^*)$; these densities solve
\begin{equation}
\begin{aligned}
    0&=\rho_A^{*}(\mu_A\rho^{*}_{\varnothing}(t) -\sigma_C\rho_C^{*} - \omega_A) \;,
    \\
    0&=\rho_B^{*}(\mu_B\rho^{*}_{\varnothing}(t) -\sigma_A\rho_A^{*} - \omega_B) \;,
    \\
    0&=\rho_C^{*}(\mu_C\rho^{*}_{\varnothing}(t) -\sigma_B\rho_B^{*} - \omega_C) \; .
\end{aligned}
\end{equation}
and none of them can be zero since that corresponds to fixation or extinction.
Hence the terms in parentheses must all vanish, which leads to $\rho^*_{\varnothing}={\cal F}$ with
\begin{equation}
\mathcal{F}=\frac{1+\frac{\omega_A}{\sigma_C}+\frac{\omega_B}{\sigma_A}+\frac{\omega_C}{\sigma_B}}{1+\frac{\mu_A}{\sigma_C}+\frac{\mu_B}{\sigma_A}+\frac{\mu_C}{\sigma_B}} . 
\end{equation}   
and therefore
\begin{equation}
\begin{aligned}
    \rho_A^{*}=\frac{\mu_B\mathcal{F}-\omega_B}{\sigma_A},
    \\
    \rho_B^{*}=\frac{\mu_C\mathcal{F}-\omega_C}{\sigma_B},
    \\
    \rho_C^{*}=\frac{\mu_A\mathcal{F}-\omega_A}{\sigma_C},
\end{aligned}
\end{equation}
Similar to the case of fixation, too large values for the death rates $\omega_A,\omega_B,\omega_C$ lead to $\rho_X^*<0$ for some species $X$ in which case the fixed point is never reached and coexistence cannot occur.

Rearranging the expression for $\rho^*_A$ yields
\begin{equation}
\begin{aligned}
    \rho^*_A &= \frac{\mu_B\mathcal{F}-\omega_B}{\sigma_A} 
    \\
    & = \frac{\mu_B\Bigl( 1+\frac{\omega_A}{\sigma_C}+\frac{\omega_B}{\sigma_A}+\frac{\omega_C}{\sigma_B} \Bigl)}{\sigma_A\Bigl( 1+\frac{\mu_A}{\sigma_C}+\frac{\mu_C}{\sigma_B} \Bigl)+\mu_B }-\frac{\omega_B}{\sigma_A}\\
    &=\frac{f(\sigma_A)}{g(\sigma_A)}-\frac{\omega_B}{\sigma_A},
\end{aligned}
\end{equation}
where $f(\sigma_A)$ is a decreasing function of $\sigma_A$ and $g(\sigma_A)$ is increasing function of $\sigma_A$.  

This establishes that the density of the smart species $\rho_A^*$ in the coexistence phase generically increases as $\sigma_A$ is reduced.  That is survival of the weakest.

\section{Effect of Perception Range $\mathcal{R}$}\label{sec:perception_range}
We briefly discuss the effect of the perception range $\mathcal{R}$ on the effectiveness of the adaptive local strategies.  In Fig.~\ref{fig:range_scan} we show the population densities as a function of local adaptive factor $\phi_{\rm{DI}}$ at  $p_{\omega}=0.015$.  The data show the effectiveness of adaptive strategies has weak dependence on the perception range.  Therefore, the main text fixes the perception range to be $\mathcal{R}=3$. We note in this section we allow $\phi$ to take values between $-1$ and $3$.

\bibliography{apssamp}

\end{document}

%% file: tab_opt.tex
\begin{tabular}{|l|l|l|}
\hline
               & sparse case, $\omega=0.015$                                                                       & crowded case, $\omega=0.005$                                                                      
\\ \hline
Symmetric (null)  \quad       & 
$(\sigma_A, \lambda_A)=
(1.0, 1.0)$                                                    
& 
$(\sigma_A, \lambda_A)=
(1.0, 1.0)$                                                    
\\ \hline
Non-directional  \quad         & 
$(\sigma_A,\lambda_A)=
(0.6,1.6)$                                                    
& 
$(\sigma_A,\lambda_A)=
(0.2, 0.4)$                                                    
\\ \hline
E          & 
$(\sigma_A,\lambda_A,\phi_{\rm{E}})
=(0.6,1.6,1.0)$
& 
$(\sigma_A,\lambda_A,\phi_{\rm{E}})
=(0.4,1.8,2.75)$                           
\\ \hline
S         & 
$(\sigma_A,\lambda_A,\phi_{\rm{S}})
=(0.6,1.6,2.75)$                           
& 
$(\sigma_A,\lambda_A,\phi_{\rm{S}})
=(0.2,0.4,1.0)$                           
\\ \hline
H           & 
$(\sigma_A,\lambda_A,\phi_{\rm{H}})
=(0.6,1.6,0.0)$                           
& 
$(\sigma_A,\lambda_A,\phi_{\rm{H}})
=(0.2,0.4,0.0)$ 
\\ \hline
C        & 
$(\sigma_A,\lambda_A,\phi_{\rm{C}})
=(0.6,1.6,0.0)$                            
& 
$(\sigma_A,\lambda_A,\phi_{\rm{C}})
=(0.2,0.4,0.75)$                         
\\ \hline
E \& S  & 
$(\sigma_A,\lambda_A,\phi_{\rm{E}},\phi_{\rm S})
=(0.6,1.6,1.25,1.75)$ 
& 
$(\sigma_A,\lambda_A,\phi_{\rm{E}},\phi_{\rm S})
=(0.4,1.8,2.75,0.0)$ 
\\ \hline
E \& H    & 
$(\sigma_A,\lambda_A,\phi_{\rm{E}},\phi_{\rm H})
=(0.6,1.6,1.00,0.0)$ 
& 
$(\sigma_A,\lambda_A,\phi_{\rm{E}},\phi_{\rm H})
=(0.4,1.8,2.75,0.0)$ 
\\ \hline
E \& C & 
$(\sigma_A,\lambda_A,\phi_{\rm{E}},\phi_{\rm C})
=(0.6,1.6,1.00,0.0)$  
& 
$(\sigma_A,\lambda_A,\phi_{\rm{E}},\phi_{\rm C})
=(0.4,1.8,2.75,0.0)$  
\\  \hline
\end{tabular}